\documentclass[aps,pra,showpacs,twocolumn,superscriptaddress]{revtex4-1}
\usepackage{amssymb}
\usepackage{amsmath}
\usepackage{graphicx}
\usepackage{epsfig}
\usepackage{subfigure}

\begin{document}

\title{Exceptional point induced lasing dynamics in a non-Hermitian system}
\author{K. L. Zhang}
\author{P. Wang}
\author{Z. Song}
\email{songtc@nankai.edu.cn}
\affiliation{School of Physics, Nankai University, Tianjin 300071, China}

\begin{abstract}
Non-Hermitian systems exhibit many peculiar dynamic behaviors which never
showed up in Hermitian systems. The existence of spectral singularity (SS)
for a non-Hermitian scattering center provides a lasing mechanism in the
context of quantum mechanics. In this paper we show exactly that a finite
system at exceptional point (EP) can also provide an alternative laser
solution. We investigate the dynamics of non-Hermitian Su-Schrieffer-Heeger
(SSH) model around EP, based on the analysis of parity-time ($\mathcal{PT}$)
and chiral-time ($\mathcal{CT}$) symmetries of the Hamiltonian. In contrast
to the SS lasing mechanism for an infinite system, such an SSH chain acts as
an active laser medium at threshold, within which a stationary particle
emission can be fired anywhere, rather than a specific location at the
non-Hermitian scattering center only. In addition, some relevant peculiar
phenomena arising from interference between wave packets are revealed based
on the analytical solutions.
\end{abstract}

\maketitle



\section{Introduction}

\label{sec.intro}

The recent development of non-Hermitian quantum mechanics \cite{Bender
98,Bender 99,Dorey 01,Dorey 02,Bender 02,A.M43,A.M36,Jones,M.Z40,M.Z41,M.Z82}
has opened up the perspective in several branches of physics \cite%
{AGuo,CERuter,Wan,Sun,LFeng,BPeng,LChang,LFengScience,HodaeiScience,NC2015}.
The remarkable features of a non-Hermitian system is the violation of
conservation law of the Dirac probability and the exceptional point (EP) or
spectral singularity (SS). Based on the former, the complex potential is
employed to describe open systems phenomenologically \cite{Muga}.
Furthermore, unconventional propagation of light associated with the
gain/loss has been demonstrated by engineering effective non-Hermitian
Hamiltonians in optical systems \cite{Guo, K. G. Makris, Z. H. Musslimani}.
On the other hand, many unique optical phenomena have been observed around
exceptional point (EP), ranging from loss-induced transparency \cite{Guo},
power oscillations violating left-right symmetry, low-power optical diodes 
\cite{PengB}, to single-mode laser \cite{FengL,Hodaei}. A fascinating
phenomenon of non-Hermitian optical systems in the application aspect is the
gain-induced detection, such as enhanced spontaneous emission \cite{LinZ},
enhanced nano-particle sensing \cite{Wiersig} as well as the amplified
transmission in the optomechanical system \cite{LiuYL,ZhangXZ}. Both
theoretical and experimental works not only give an insight into the
dynamical property of the non-Hermitian Hamiltonian but also provide a
platform to implement the optical phenomenon.

\begin{figure}[tbp]
\centering
\includegraphics[ bb=0 123 550 657, width=0.5\textwidth, clip]{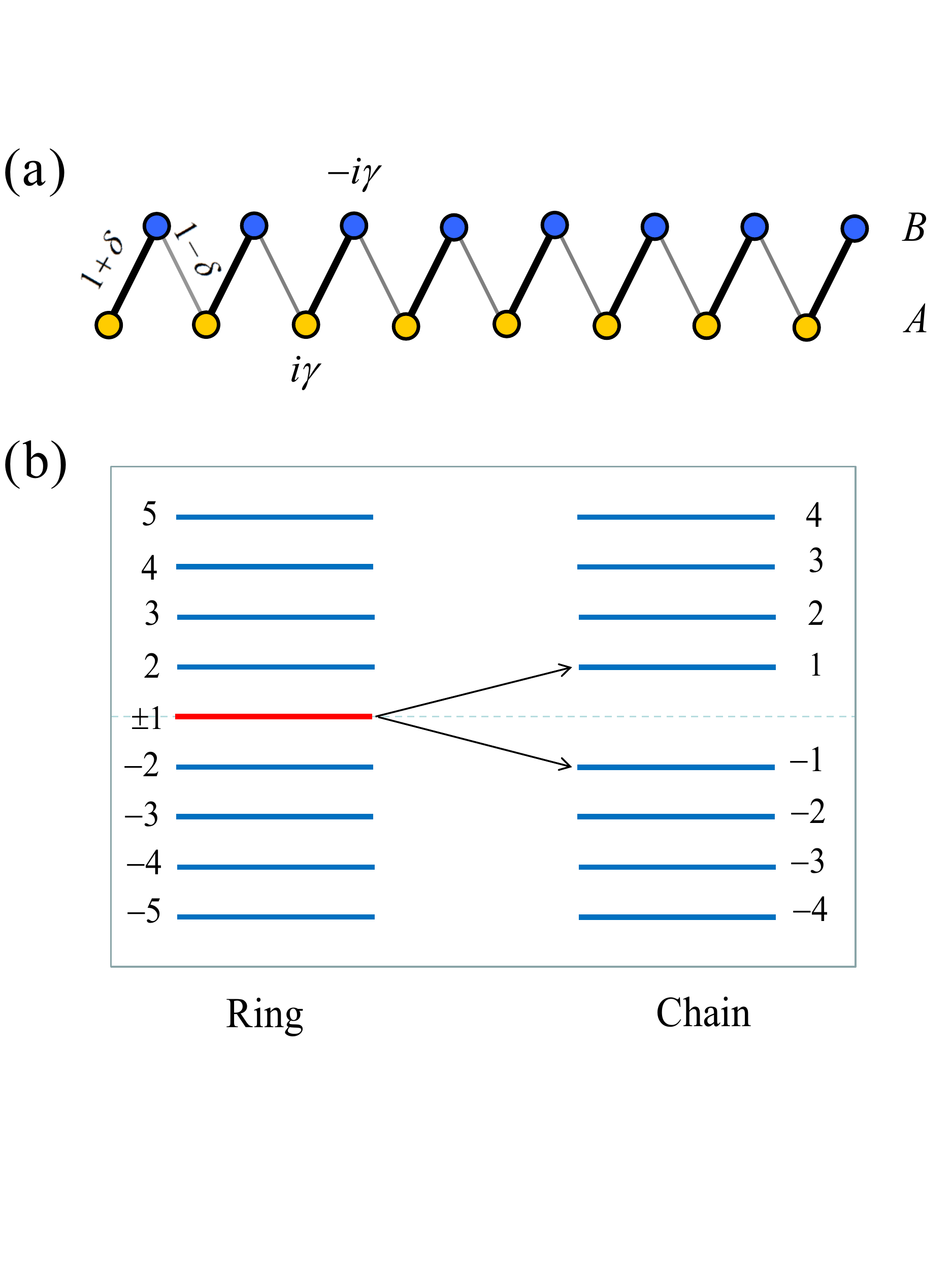} 
\caption{(Color online) (a) Schematic for $1$D non-Hermitian SSH model. It
consists of two sublattice, $A$ (golden) and $B$ (blue). Black thick and
gray thin lines indicate the hopping between two nearest neighbor sites with
amplitudes $\left( 1+\protect\delta \right) $\ and $\left( 1-\protect\delta \right) $, respectively. Two sublattices have opposite site imaginary
potentials $\pm i\protect\gamma $, representing the physical gain and loss.
It has $\mathcal{PT}$-symmetry and can have full real spectrum in the case
of $1>\protect\delta >0$\ and $\protect\gamma <\protect\gamma _{c}$\ (see
text). (b) The spectra of the non-Hermitian SSH model at $\protect\gamma =\protect\gamma _{c}$\ with periodic and open boundary (by breaking one of
the weak hopping term) conditions, respectively. In both two cases, energy
levels near zero are equal-spacing. For ring system two zero level
coalescence which is splitted into two levels when the open boundary
condition is imposed. } \label{Fig1}
\end{figure}

In this paper, using the exact solutions, we introduce a mechanism for
self-sustained emission in finite non-Hermitian systems at the EP. We show
that, without the existence of SS in a scattering system, one is able to
obtain a class of lasing solutions. For such solutions, a lasing mode is
fired at any location of the system, which acts as an active lasing medium,
rather than the non-Hermitian scattering center only in the context of SS
regime. In order to demonstrate the EP lasing dynamics, we consider a one-dimensional (1D) $\mathcal{PT}$-symmetry non-Hermitian Su-Schrieffer-Heeger (SSH) model \cite%
{HWH}. The exact solution is obtained in the strong dimerization limit,
which shows that the equal-level-spacing high-frequency standing-wave modes
(EHSM) \cite{ZKL} can be achieved in a chain system when the corresponding
ring system is tuned at the exceptional point (EP), as depicted
schematically in Fig. \ref{Fig1}. Such a deliberately designed system
supports some peculiar dynamics for a localized initial state, which
originates from the combination of the time evolutions involving nonzero
energy levels and Jordan block of the corresponding SSH ring lattice within
a finite time scale. Based on the analysis of parity-time ($\mathcal{PT}$)
and chiral-time ($\mathcal{CT}$) symmetries of the Hamiltonian, a class of
solutions are constructed based on energy levels around the zero energy.
Such a class of states involve wave packets with different shapes and
locations, which support the stationary lasing dynamics. In addition, it is
found that the superposition of these states exhibit some counterintuitive
dynamical behaviors. Although the system is non-interacting, a delicate
design of the interference process results in the phenomena of
wave packet-pair annihilation and creation.

The remainder of this paper is organized as follows. In Sec. \ref{Model and
solution}, we present a non-Hermitian\ SSH chain model and the formulation
of approximate diagonalization. In Sec. \ref{Quasi-symmetric dynamics}, we
investigate the generic dynamics for a system with parity-time ($\mathcal{PT}
$) and chiral-time ($\mathcal{CT}$) symmetries Hamiltonian. Section \ref%
{Lasing dynamics} presents the laser solution and reveals the lasing
dynamics for specific initial localized states. Section \ref{Probability
preserving and elastic collision} demonstrates some peculiar dynamical
behaviors in the present system. Finally, we give a summary and discussion
in Sec. \ref{Summary}.

\section{Model and solution}

\label{Model and solution}

As beginning, we briefly summarize known properties of the 1D non-Hermitian
SSH model with staggered balanced gain and loss. It has been studied
systematically in the previous work \cite{ZKL, HWH}. The simplest
tight-binding model with these features is 
\begin{eqnarray}
H &=&(1+\delta )\sum_{j=1}^{N}a_{j}^{\dag }b_{j}+(1-\delta )\sum_{j=1}^{N-1%
\text{ or }N}b_{j}^{\dag }a_{j+1}+\mathrm{H.c.}  \notag \\
&&+i\gamma \sum_{j=1}^{N}(a_{j}^{\dag }a_{j}-b_{j}^{\dag }b_{j}),  \label{H}
\end{eqnarray}%
where $\delta $ and $i\gamma $, are the distortion factor with unit
tunneling constant and the alternating imaginary potential magnitude,
respectively. Here $a_{l}^{\dag }$ and $b_{l}^{\dag }$ are the creation
operator of the particle at the $l$th site in $A$ and $B$\ sublattices. The
particle can be fermion or boson, depending on their own commutation
relations. In the second term, two kinds of summation can be taken,
correspond to open and periodic boundary conditions. For nonzero $\gamma $,
it is still a $\mathcal{PT}$-symmetry. Here, the time reversal operation $%
\mathcal{T}$ is such that $\mathcal{T}i\mathcal{T}=-i$, while the effect of
the parity is such that $\mathcal{P}a_{l}\mathcal{P}=b_{N+1-l}$ and $%
\mathcal{P}b_{l}\mathcal{P}=a_{N+1-l}$. Applying operators $\mathcal{P}$ and 
$\mathcal{T}$ on the Hamiltonian (\ref{H}), one has $\left[ \mathcal{T},H%
\right] \neq 0$ and $\left[ \mathcal{P},H\right] \neq 0$, but 
\begin{equation}
\left[ \mathcal{PT},H\right] =0.  \label{PT H}
\end{equation}%
In parallel, $H$ also has chiral-time symmetry 
\begin{equation}
\left\{ \mathcal{CT},H\right\} =0,  \label{CT H}
\end{equation}%
where operator $\mathcal{C}$\ is defined as

\begin{equation}
\mathcal{C}a_{j}\mathcal{C}^{-1}=a_{j},\mathcal{C}b_{j}\mathcal{C}%
^{-1}=-b_{j}.  \label{C}
\end{equation}%
The situation here is a little different from the case associated with $%
\mathcal{PT}$ symmetry. In quantum mechanics, we say that a Hamiltonian $H$\
has a symmetry represented by a operator $\mathcal{L}$\ if $[H,\mathcal{L}%
]=0 $. The word \textquotedblleft symmetry\textquotedblright\ is also used
in a different sense in condensed matter physics. We say that a system with
Hamiltonian $H$\ has chiral symmetry, if $\{H,\mathcal{C}\}=0$. The physics
of $\mathcal{C}$\ depends on the model discussed \cite%
{Asboth2016lnp,Malzard2015prl,Guo2015prb,ChenS2015prb,Peng2016np,Lee2016prl}%
. A non-Hermitian system with $\mathcal{CT}$\ symmetry has been
systematically studied in Ref. \cite{LS17 PRA}.

According to the non-Hermitian quantum theory, such a Hamiltonian may have
fully real spectrum within a certain parameter region. The boundary of the
region is the critical point of quantum phase transition associated with $%
\mathcal{PT}$-symmetry breaking. For the system with periodic boundary
condition, the critical point occurs at $\gamma =\gamma _{c}=2\delta $,
which is also referred as to EP \cite{HWH}. According to the Appendix in
Ref. \cite{ZKL}, in the strong dimerization limit $1+\delta \gg 1-\delta $,
the Hamiltonian with open boundary condition can be diagonalized
approximately and the single-particle eigen vectors for $\gamma =\gamma _{c}$%
\ can be expressed as%
\begin{eqnarray}
\left\vert \psi _{n}^{\pm }\right\rangle &=&\sqrt{\frac{\pm \left( -1\right)
^{N}}{N+1}}\sum_{j=1}^{N}(-1)^{j}\sin \left( kj\right)  \notag \\
&&\times \left( e^{\pm i\varphi _{k}/2}a_{j}^{\dag }\pm e^{\mp i\varphi
_{k}/2}b_{j}^{\dag }\right) \left\vert 0\right\rangle ,
\end{eqnarray}%
\ where $k$ is defined as%
\begin{equation}
k=\frac{n\pi }{N+1},\text{ }n\in \lbrack 1,N].
\end{equation}%
The corresponding eigen energy is%
\begin{equation}
\varepsilon _{k}=\sqrt{[(1+\delta )-(1-\delta )\cos k]^{2}-\gamma _{c}^{2}},
\end{equation}%
and%
\begin{equation}
\tan \varphi _{k}=\frac{\gamma _{c}}{\varepsilon _{k}}.
\end{equation}%
We note that $\gamma _{c}$\ is no longer the EP for the open chain and we
find that the energy levels can be expressed as 
\begin{equation}
E_{n}^{\pm }=\pm n\omega ,\text{ }\omega =\frac{\sqrt{2\delta (1-\delta )}%
\pi }{N+1},
\end{equation}%
approximately for not large $n$. Here $\left\vert \psi _{n}^{\pm
}\right\rangle $\ is normalized in the framework of Dirac inner product,
i.e., $\langle \psi _{m}^{\pm }\left\vert \psi _{n}^{\pm }\right\rangle
=\delta _{mn}$. It indicates that the spectrum $\varepsilon _{k}$ consists
of two branches separated by an energy gap $\Delta =2\omega $, which ensures
the existence of the equal-level-spacing standing-wave modes (ESM) around
zero energy. In this model, the value of $\gamma _{c}$ is necessary for
achieving a set of eigen states as ESM, which is crucial for the
construction of our target state. In addition, the deliberate expression of
of the set of eigenvector $\left\{ \left\vert \psi _{n}^{\pm }\right\rangle
\right\} $\ makes it satisfies%
\begin{equation}
\mathcal{PT}\left\vert \psi _{n}^{\pm }\right\rangle =(-1)^{n}\left\vert
\psi _{n}^{\pm }\right\rangle ,  \label{PT-n}
\end{equation}%
and%
\begin{equation}
\mathcal{CT}\left\vert \psi _{n}^{\pm }\right\rangle =-i\left\vert \psi
_{n}^{\mp }\right\rangle ,  \label{CT-n}
\end{equation}%
which will be used to analyze the dynamics of the system in the following
sections. It is familiar to the operator $\mathcal{P}$\ or $\mathcal{C}$\ in
a system with refection symmetry, which is widely used in physics. However,
we would like to point out that here $\mathcal{T}$\ is an antilinear operator,
which has not a definite eigenvalue for a given eigenstate. In the previous
work \cite{ZKL}, we have studied a similar model and the corresponding
dynamics. The spectrum consists of two branches separated by an energy gap $%
\Delta \succsim 0$\ and then it may or not contain the coalescing level with
zero energy. In addition, since only one branch of levels are involved in
the initial states, the obtained result is not sensitive to the precise
value of the gap $\Delta $. However, in the present work we will show that
the existence of stationary laser mode strongly depends on the value of $%
\Delta $, or $\gamma $.

\section{Quasi-symmetric dynamics}

\label{Quasi-symmetric dynamics} Before the investigation for the dynamics
of some specific initial state, we would like to study some general
properties of dynamics\ for a $\mathcal{PT}$ and $\mathcal{CT}$\ symmetric
system, which is helpful for the subsequent discussions. In the following
two sections, we will reveal two features of dynamics, which are related to
the symmetries of the system, but not exact results. The first one is a
quasi symmetric time evolution and the second one is about time-reflection
symmetry. The obtained result in this section is applicable for more general
system.

Unlike the symmetry related to a linear operator $\mathcal{L}$, here $%
\mathcal{PT}$ and $\mathcal{CT}$\ are anti-linear operators, which act in a
different way in the time evolution for a symmetric initial state. Consider
a Hamiltonian $\mathcal{H}$\ has $\mathcal{L}$, $\mathcal{PT}$ and $\mathcal{%
CT}$\ symmetries, i.e., 
\begin{equation}
\left[ \mathcal{L}\text{,}\ \mathcal{H}\right] =\left[ \mathcal{PT}\text{,}\ 
\mathcal{H}\right] =\left\{ \mathcal{CT}\text{,}\ \mathcal{H}\right\} =0,
\end{equation}%
respectively. An initial state $\left\vert \psi \left( 0\right)
\right\rangle $ is taken as the eigen state of $\mathcal{L}$, $\mathcal{PT}$
and $\mathcal{CT}$, e.g., 
\begin{equation}
\mathcal{L}\left\vert \psi \left( 0\right) \right\rangle =\mathcal{PT}%
\left\vert \psi \left( 0\right) \right\rangle =\mathcal{CT}\left\vert \psi
\left( 0\right) \right\rangle =\left\vert \psi \left( 0\right) \right\rangle
.
\end{equation}%
We are interested in the symmetry of the evolved state $\left\vert \psi
\left( t\right) \right\rangle =e^{-i\mathcal{H}t}\left\vert \psi \left(
0\right) \right\rangle $. Direct derivations show that%
\begin{eqnarray}
\mathcal{L}\left\vert \psi \left( t\right) \right\rangle &=&\mathcal{CT}%
\left\vert \psi \left( t\right) \right\rangle =\left\vert \psi \left(
t\right) \right\rangle ,  \label{L CT} \\
\mathcal{PT}\left\vert \psi \left( t\right) \right\rangle &=&\left\vert \psi
\left( -t\right) \right\rangle ,  \label{PT t}
\end{eqnarray}%
which indicate that the evolved state may not maintain the initial symmetry
associated with an anti-linear operator.

Applying the above analysis to the present SSH model, we will have following
observations. Here $\mathcal{L}$\ can be taken as operator $%
\sum_{j=1}^{N}(a_{j}^{\dag }a_{j}+b_{j}^{\dag }b_{j})$\ as an example, which
obeys the Eq. (\ref{L CT}). For the $\mathcal{PT}$\ symmetric initial state,
we have the symmetric Dirac probability distribution 
\begin{eqnarray}
\left\vert \langle j\left\vert \psi \left( 0\right) \right\rangle
\right\vert ^{2} &=&\left\vert \left\langle j\right\vert \mathcal{PT}%
\left\vert \psi \left( 0\right) \right\rangle \right\vert ^{2}  \notag \\
&=&\left\vert \langle 2N+1-j\left\vert \psi \left( 0\right) \right\rangle
\right\vert ^{2}.
\end{eqnarray}%
However, for the evolved sate we have%
\begin{eqnarray}
\left\vert \langle j\left\vert \psi \left( t\right) \right\rangle
\right\vert ^{2} &=&\left\vert \left\langle j\right\vert \left( \mathcal{PT}%
\right) ^{-1}e^{iHt}\mathcal{PT}\left\vert \psi \left( 0\right)
\right\rangle \right\vert ^{2}  \notag \\
&=&\left\vert \left\langle 2N+1-j\right\vert \psi \left( -t\right) \rangle
\right\vert ^{2},  \label{t -t}
\end{eqnarray}%
which cannot guarantee a symmetric probability profile.

Now we will show that the time evolution is symmetric approximately for a
class of initial state. For small $n$, the approximate eigen state of the
Hamiltonian with even $N$ reads 
\begin{eqnarray}
\left\vert \psi _{n}^{\pm }\right\rangle &\approx &e^{\pm i\pi /4}\sqrt{%
\frac{\pm 1}{N+1}}\sum_{j=1}^{N}(-1)^{j}\sin \left( kj\right)  \notag \\
&&\times \left( a_{j}^{\dag }-ib_{j}^{\dag }\right) \left\vert
0\right\rangle ,
\end{eqnarray}%
by taking%
\begin{equation}
\varphi _{k}\approx \frac{\pi }{2}.
\end{equation}%
The implication of the approximation is clear that two eigen state $%
\left\vert \psi _{n}^{\pm }\right\rangle $\ with opposite energy $\pm E_{n}$%
\ have the same expression. The time evolution of $\left\vert \psi _{n}^{\pm
}\right\rangle $\ is 
\begin{eqnarray}
e^{-iHt}\left\vert \psi _{n}^{\pm }\right\rangle &\approx &e^{\mp
iE_{n}t}e^{\pm i\pi /4}\sqrt{\frac{\pm 1}{N+1}}\sum_{j=1}^{N}(-1)^{j}  \notag
\\
&&\times \sin \left( kj\right) \left( a_{j}^{\dag }-ib_{j}^{\dag }\right)
\left\vert 0\right\rangle .
\end{eqnarray}%
Furthermore, for a $\mathcal{CT}$\ symmetric state, e.g.%
\begin{equation}
\left\vert \varphi _{+}\right\rangle =\sum_{n}c_{n}^{+}\left( \left\vert
\psi _{n}^{+}\right\rangle +\left\vert \psi _{n}^{-}\right\rangle \right) ,
\end{equation}%
with real $c_{n}^{+}$, we have%
\begin{eqnarray}
e^{-iHt}\left\vert \varphi _{+}\right\rangle &=&\sum_{n=1}c_{n}^{+}\left(
e^{-iE_{n}t}\left\vert \psi _{n}^{+}\right\rangle +e^{iE_{n}t}\left\vert
\psi _{n}^{-}\right\rangle \right)  \notag \\
&\approx &\sqrt{\frac{1}{N+1}}\sum_{j=1}^{N}f(j,t)\left( a_{j}^{\dag
}-ib_{j}^{\dag }\right) \left\vert 0\right\rangle ,
\end{eqnarray}%
where%
\begin{equation}
f(j,t)=\sum_{n}(-1)^{j}c_{n}^{+}\left( e^{-iE_{n}t}e^{i\pi /4}+i\text{c.c.}%
\right) \sin \left( kj\right) .
\end{equation}%
Function $\sin \left( kj\right) $ is odd (even) function about the center of
the chain when $n$ is even (odd). If the $\left\vert \varphi
_{+}\right\rangle $ is a $\mathcal{PT}$ symmetric state, the summation in $%
f(j,t)$\ runs over even (or odd) $n$ only. Function $f(j,t)$\ is also
symmetric due to the fact that any combination of odd (even) functions is
also an odd (even) function. Then probability $\left\vert \left\langle
0\right\vert \left( a_{j}+ib_{j}\right) e^{-iHt}\left\vert \varphi
_{+}\right\rangle \right\vert ^{2}$ is a symmetric function, i.e.,%
\begin{equation}
\left\vert \langle j\left\vert \psi \left( t\right) \right\rangle
\right\vert ^{2}\approx \left\vert \left\langle 2N+1-j\right\vert \psi
\left( t\right) \rangle \right\vert ^{2}.  \label{t t}
\end{equation}%
Together with Eq. (\ref{t -t}), we have%
\begin{equation}
\left\vert \langle j\left\vert \psi \left( t\right) \right\rangle
\right\vert ^{2}\approx \left\vert \langle j\left\vert \psi \left( -t\right)
\right\rangle \right\vert ^{2},  \label{time-reflection}
\end{equation}%
which indicates that the time evolution has time-reflection symmetry about
zero $t$.

We conclude that, the evolved state has symmetric probability distribution
and time-reflection symmetry\ if the initial state satisfies three
conditions, (i) $\mathcal{PT}$ symmetry, (ii) $\mathcal{CT}$ symmetry, (iii)
involving very small $n$. We would like to point out that these results are
approximate rather than exact, which are referred as to quasi symmetric
dynamics. This result is important to construct and characterize the laser
mode in the present non-Hermitian SSH chain.

\begin{figure}[b]
\centering
\includegraphics[width=0.5\textwidth, clip]{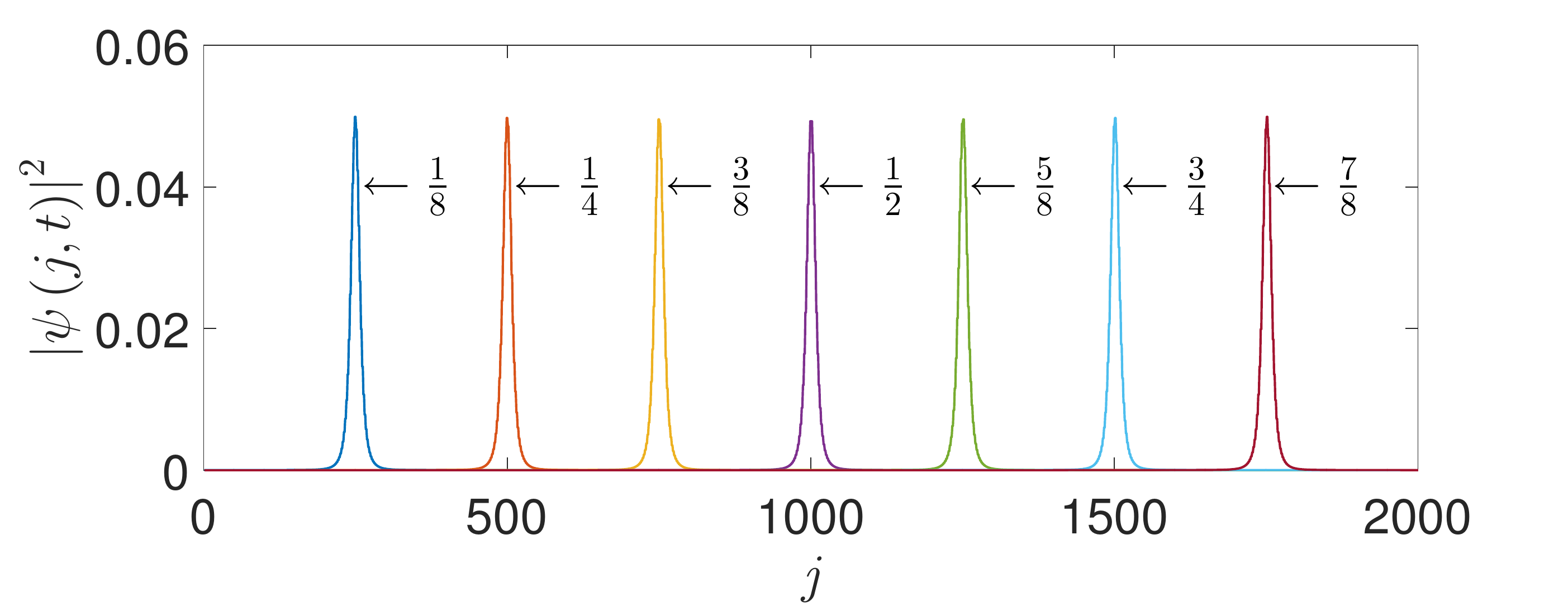} 
\caption{(Color online) Profiles of initial states from the plot of Eq. (\protect\ref{initial_state}) with $\protect\kappa _{0}=\protect\pi /8$, $\protect\pi /4$, ..., $7\protect\pi /8$, respectively. It shows that the
initial states are all localized with the identical shape in coordinate
space and the central position is the linear function of $\protect\kappa _{0} $. The parameters are $2N=2000$, $\protect\delta =0.9$, $\protect\gamma =1.8$ and $q=0.02$.}
\label{Fig2}
\end{figure}

\begin{figure*}[t]
\centering
\subfigure{\includegraphics[height=1.45in,width=1.7in]{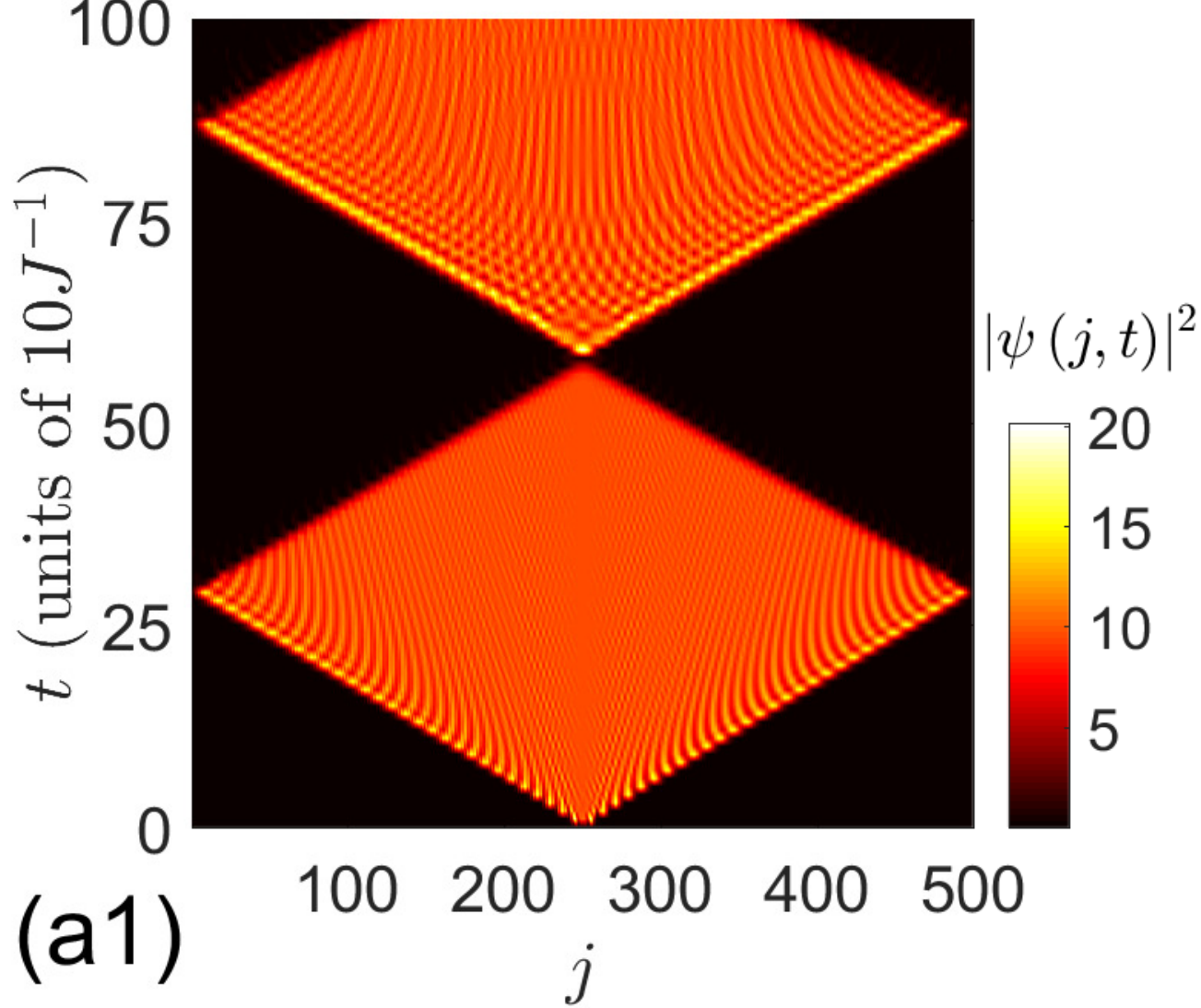}} %
\subfigure{\includegraphics[height=1.55in,width=1.65in]{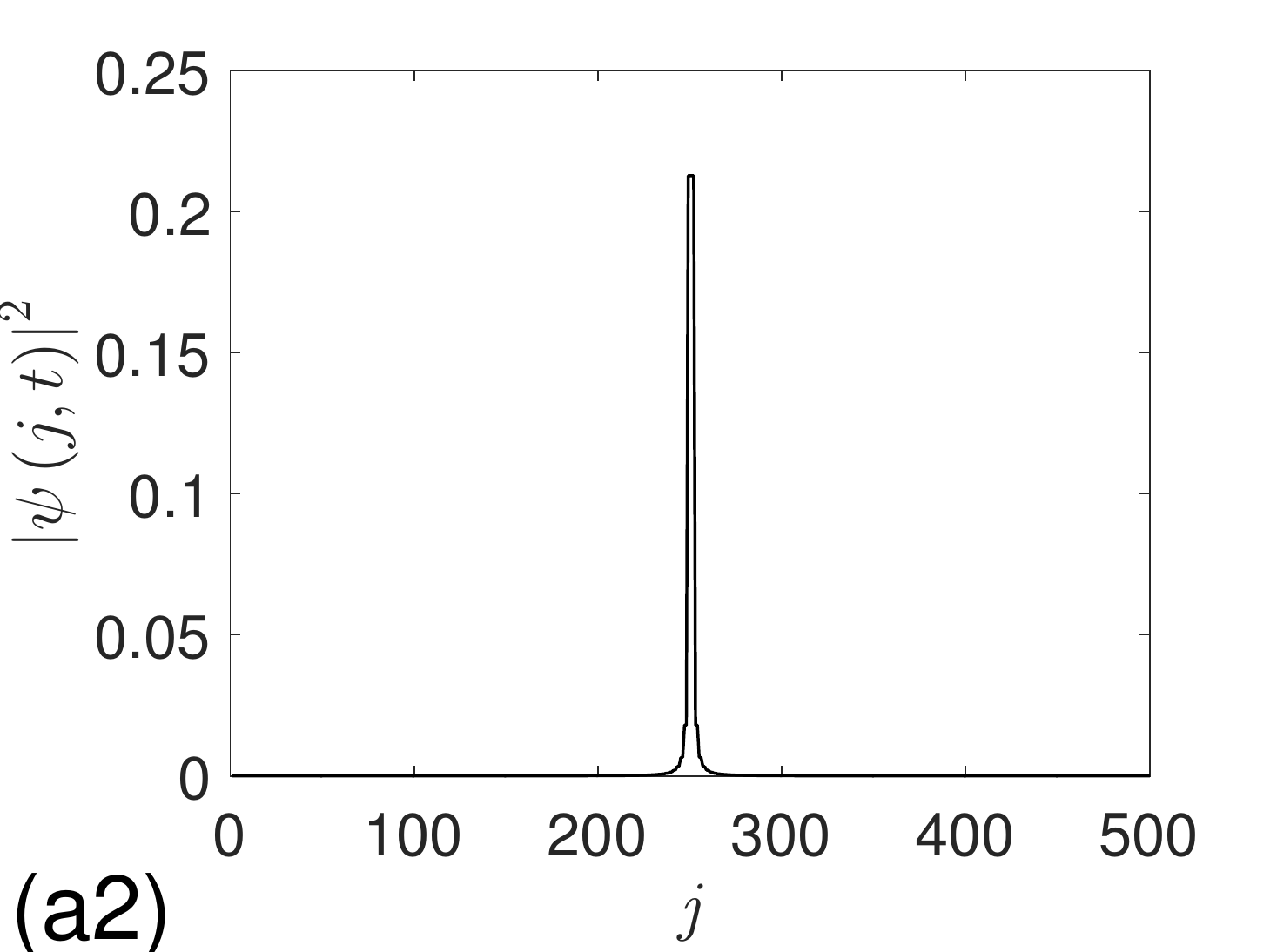}} %
\subfigure{\includegraphics[height=1.55in,width=1.76in]{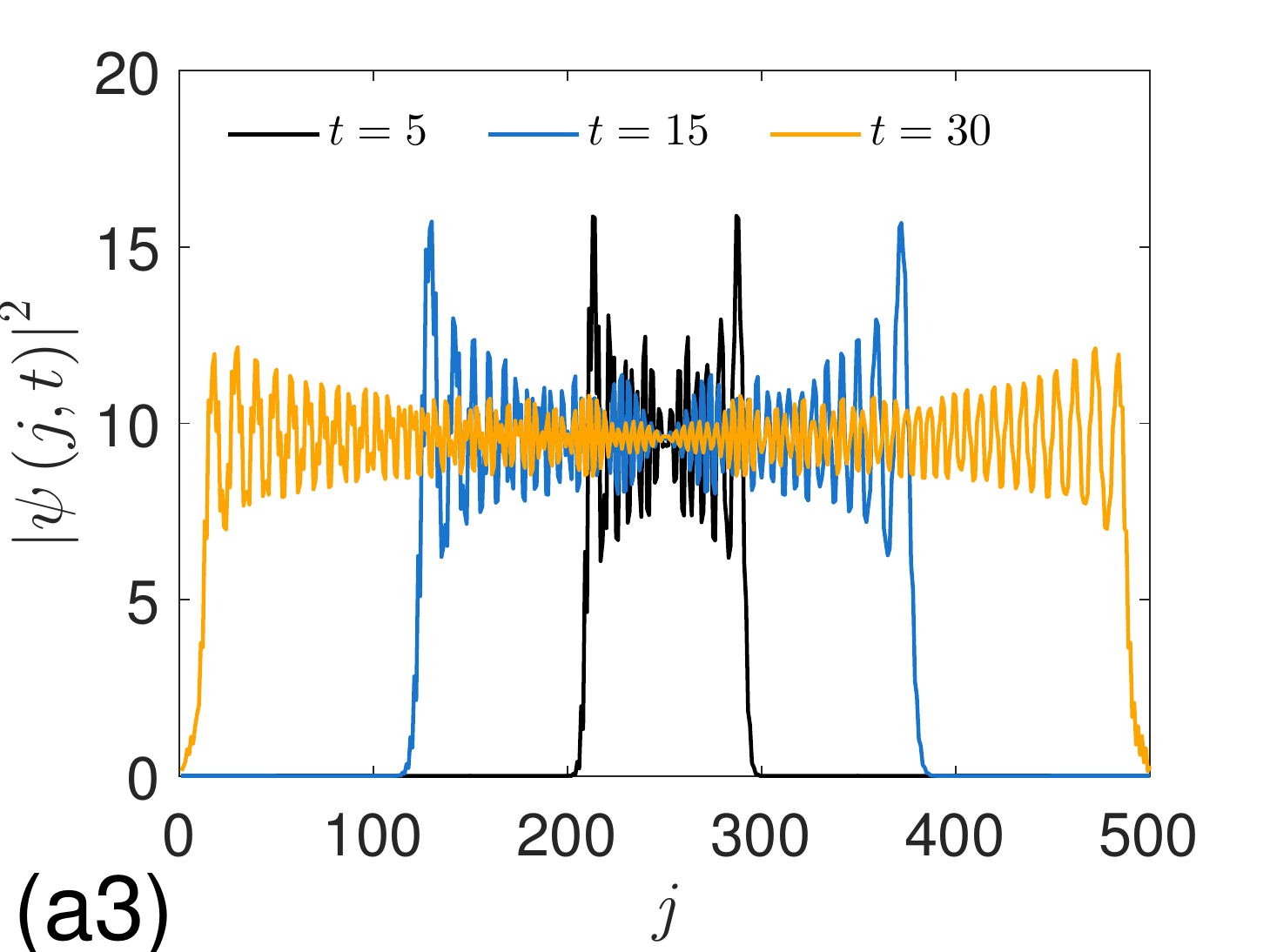}} %
\subfigure{\includegraphics[height=1.55in,width=1.76in]{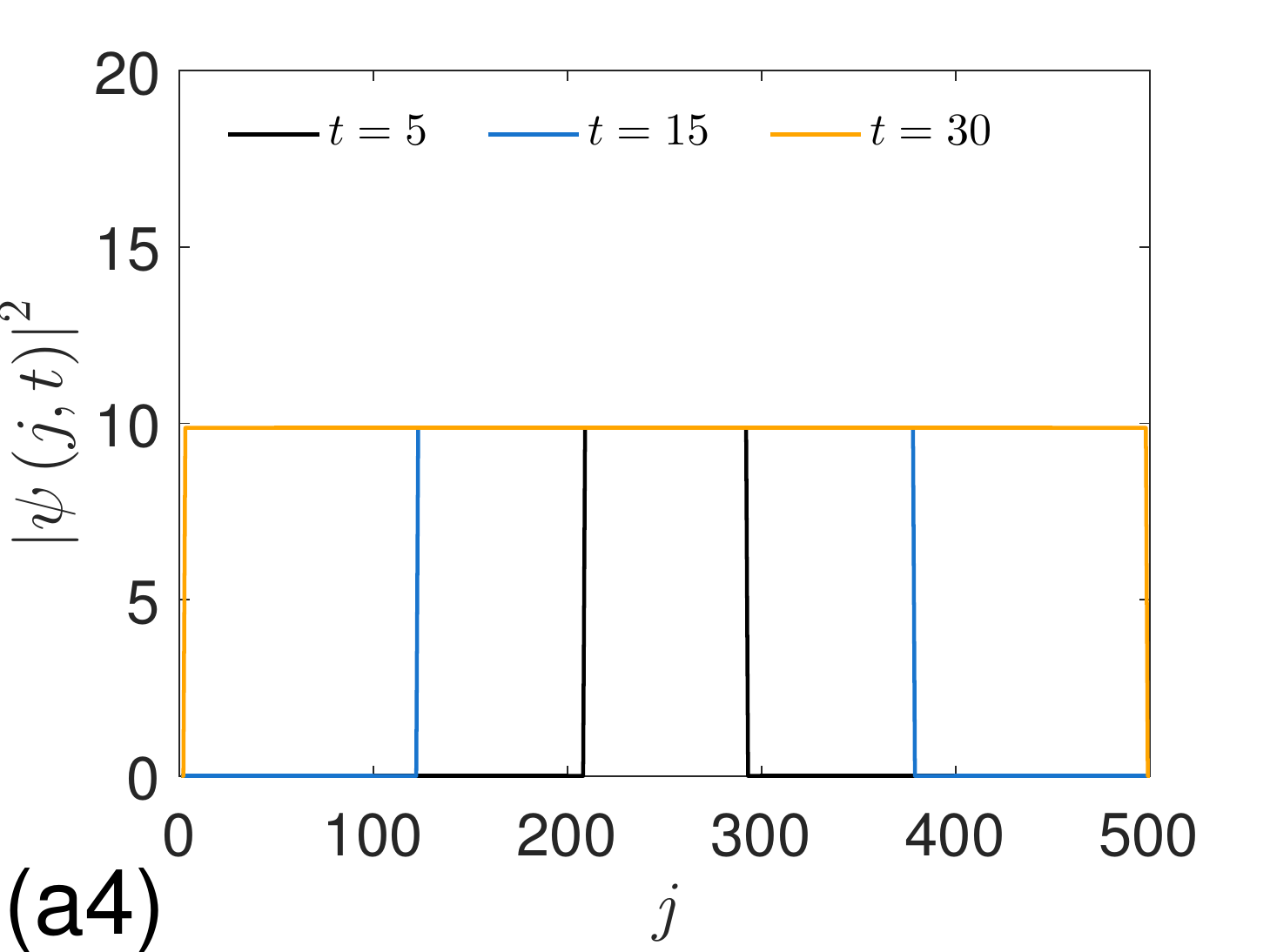}}
\par
\subfigure{\includegraphics[height=1.45in,width=1.7in]{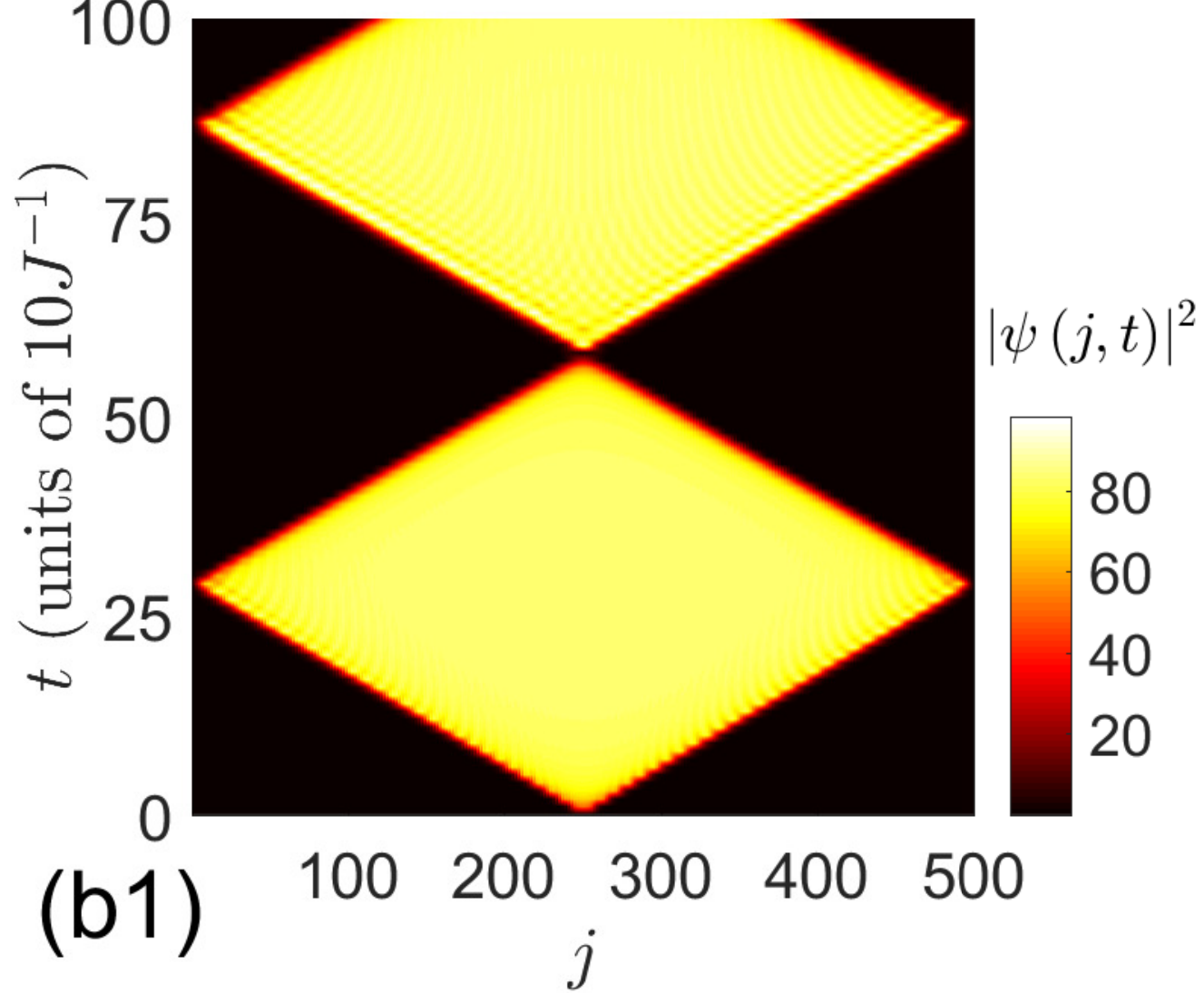}} %
\subfigure{\includegraphics[height=1.55in,width=1.65in]{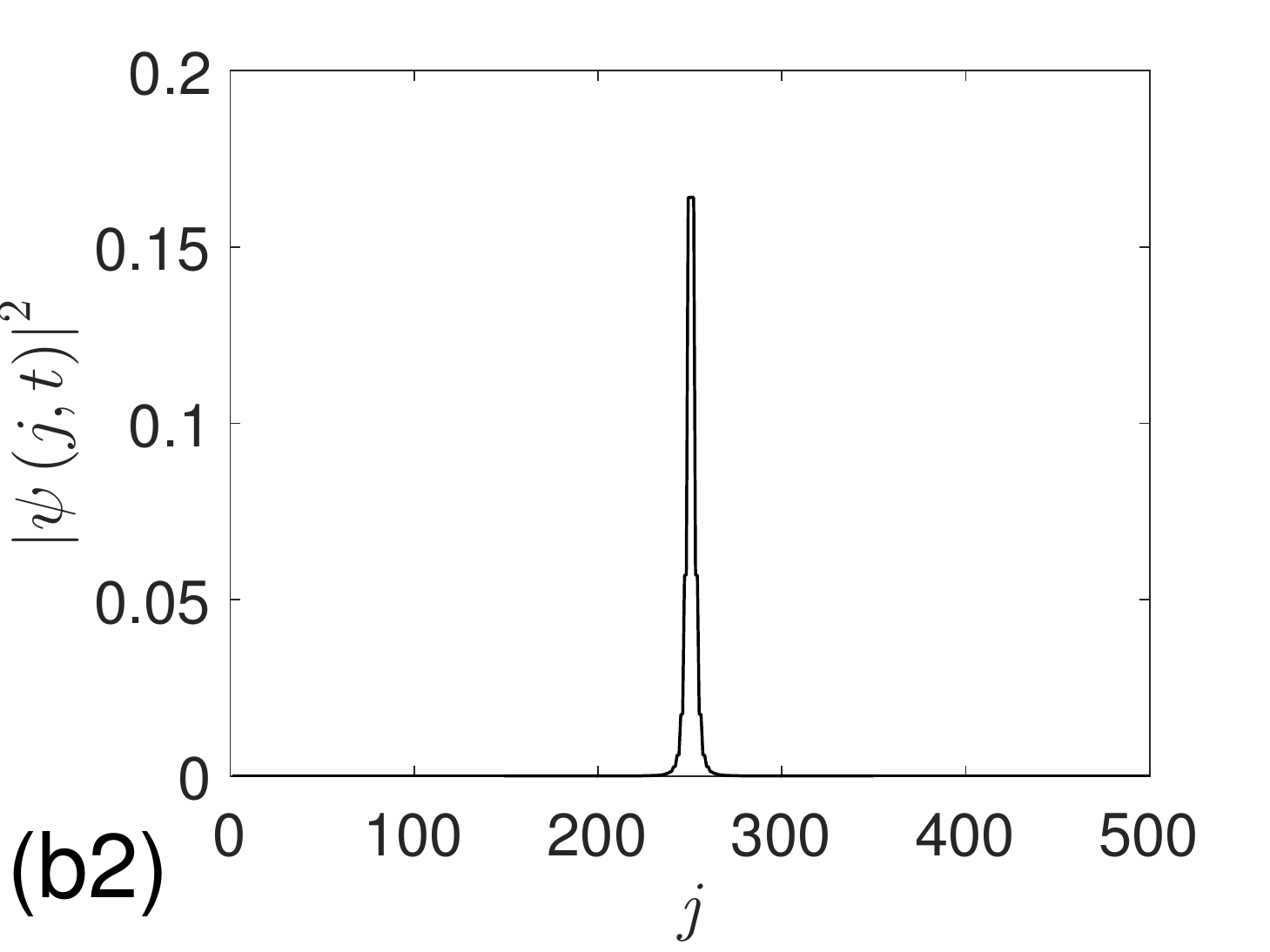}} %
\subfigure{\includegraphics[height=1.55in,width=1.76in]{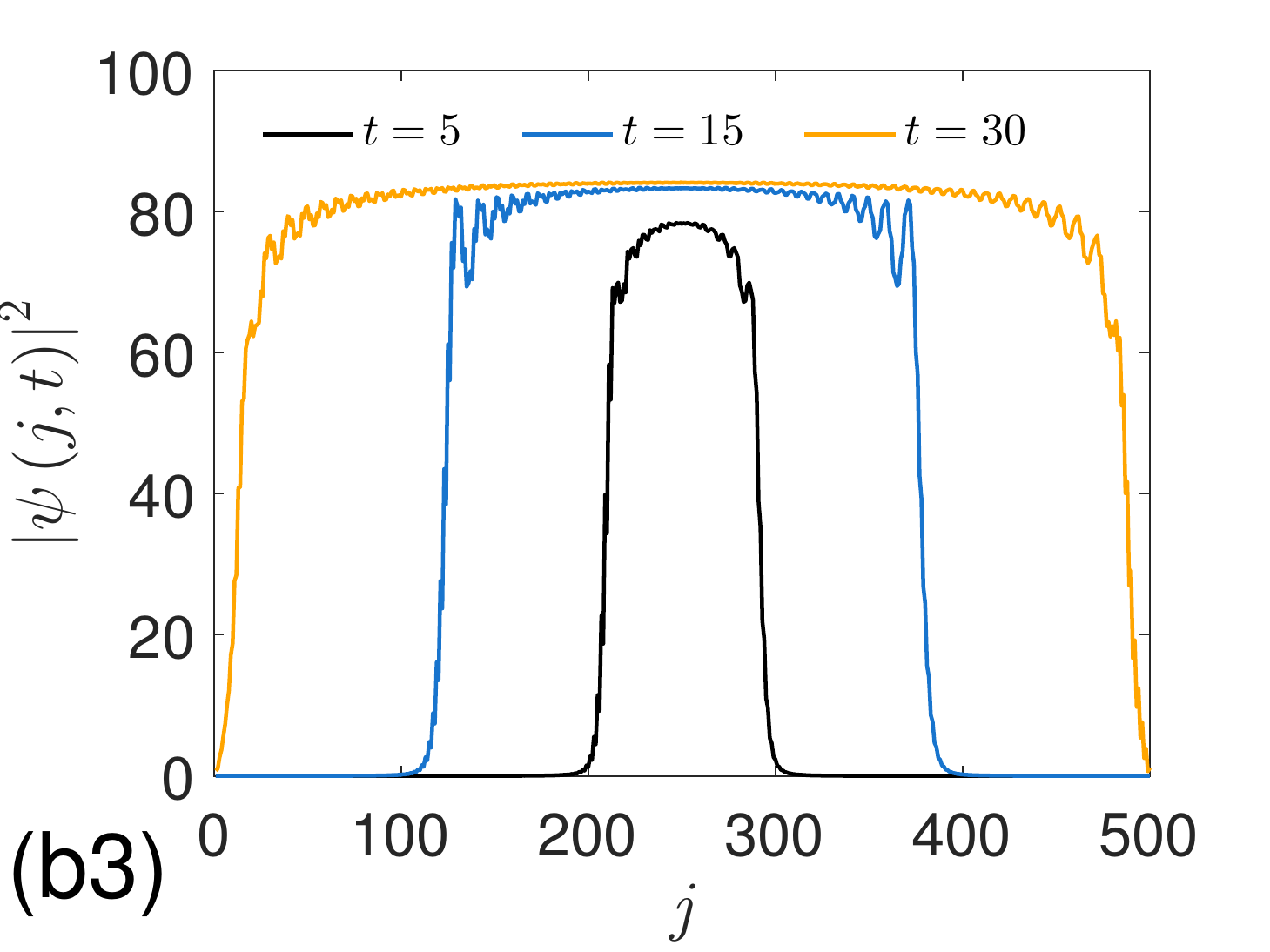}} %
\subfigure{\includegraphics[height=1.55in,width=1.76in]{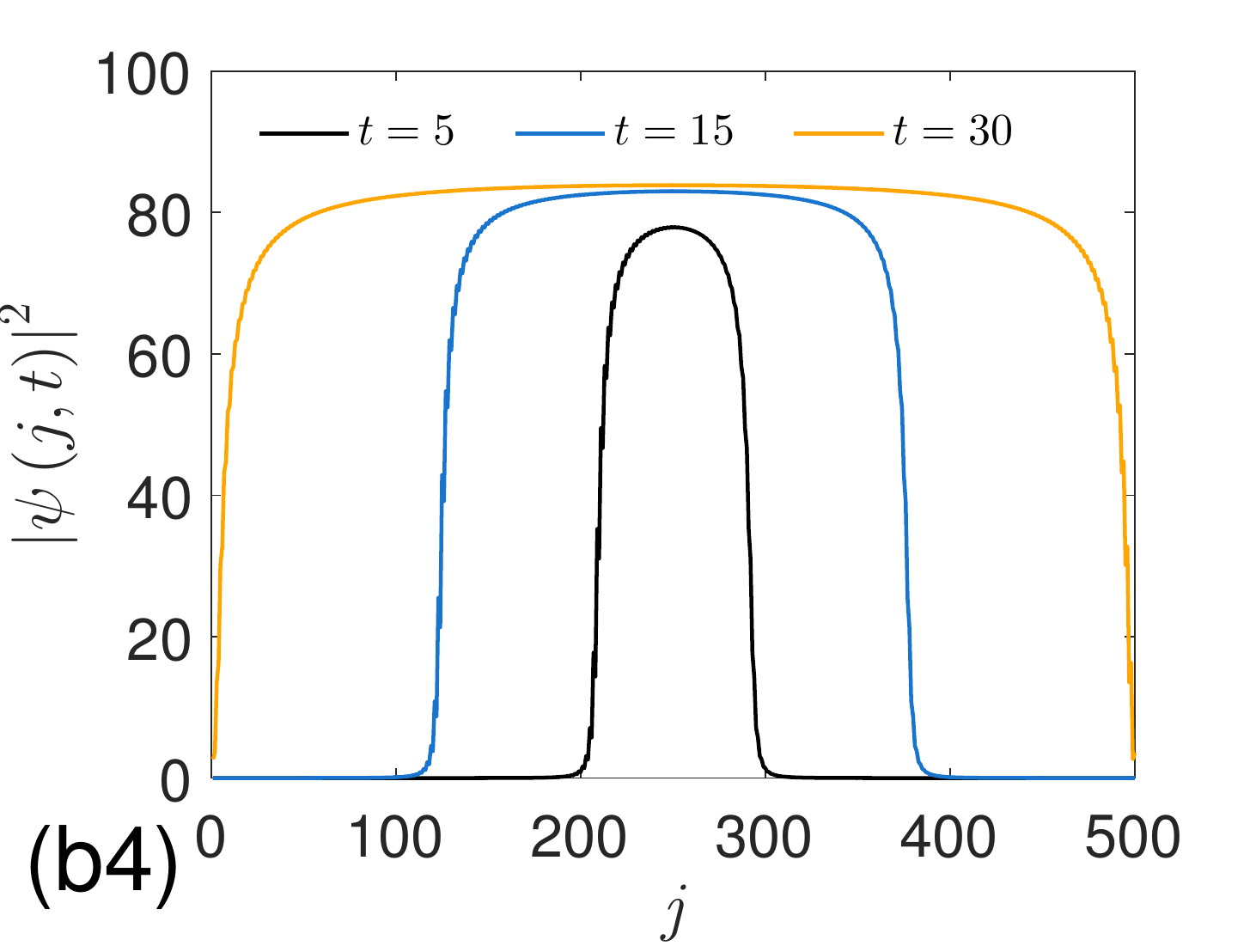}}
\par
\subfigure{\includegraphics[height=1.45in,width=1.7in]{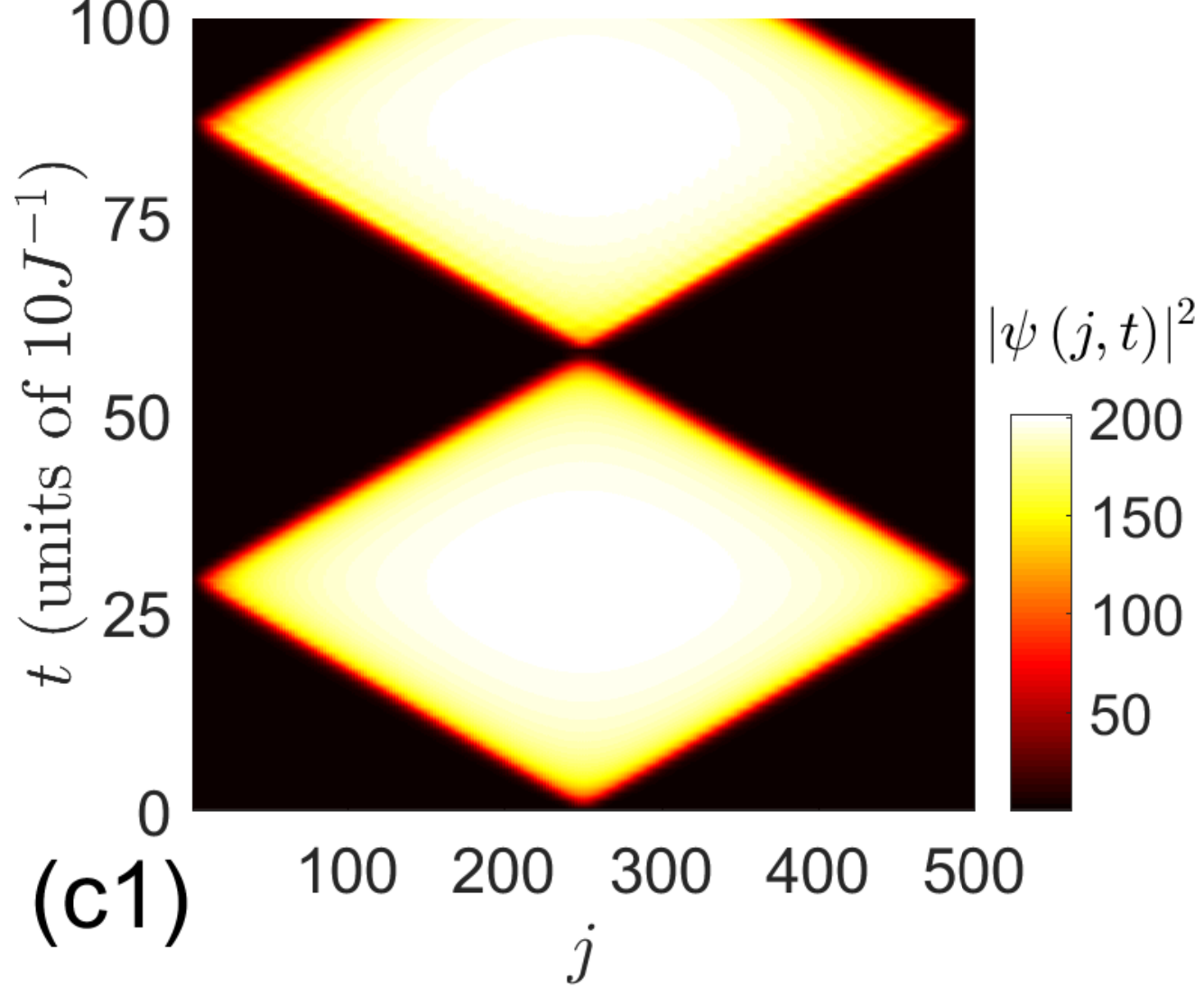}} %
\subfigure{\includegraphics[height=1.55in,width=1.65in]{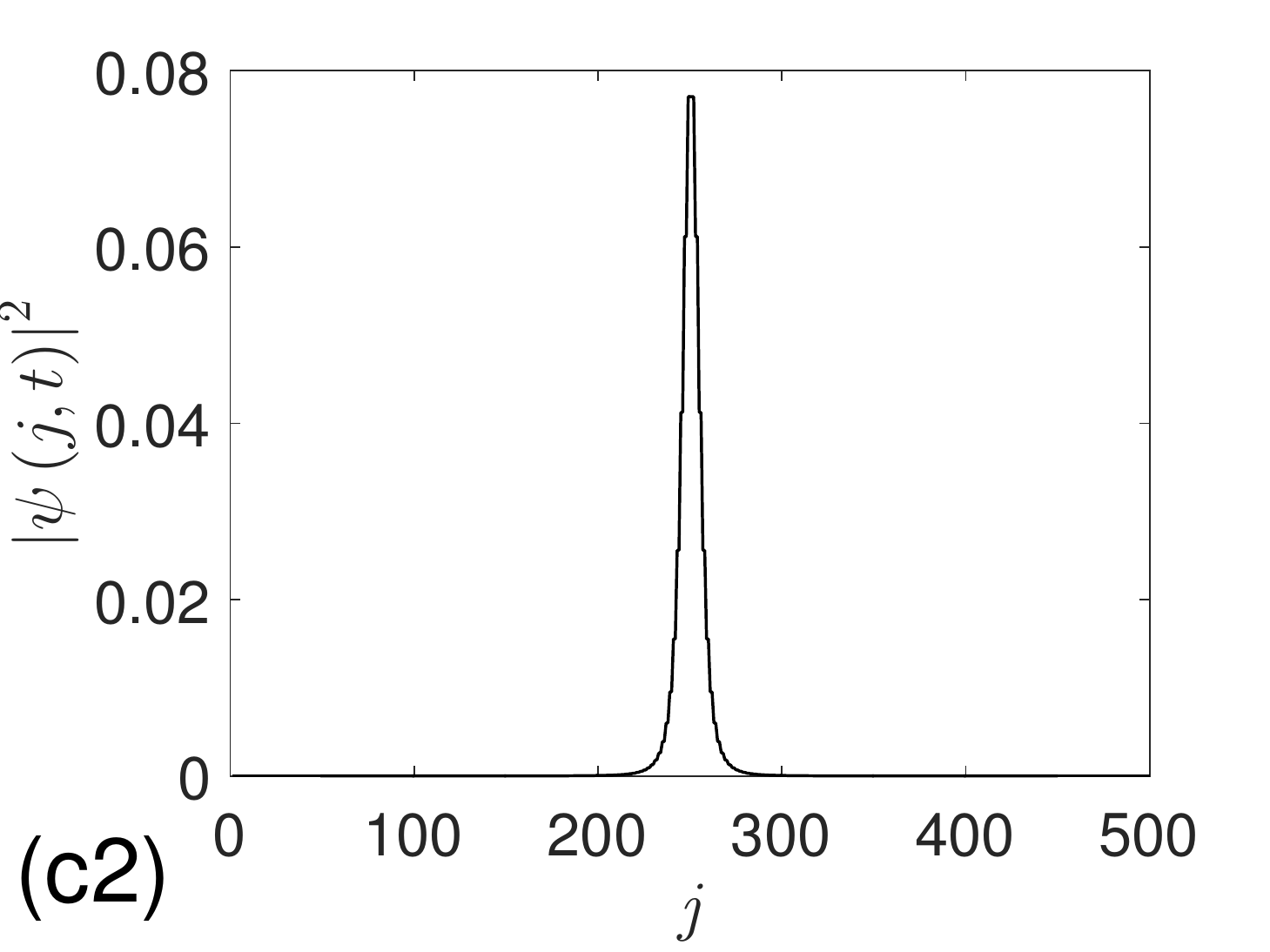}} %
\subfigure{\includegraphics[height=1.55in,width=1.76in]{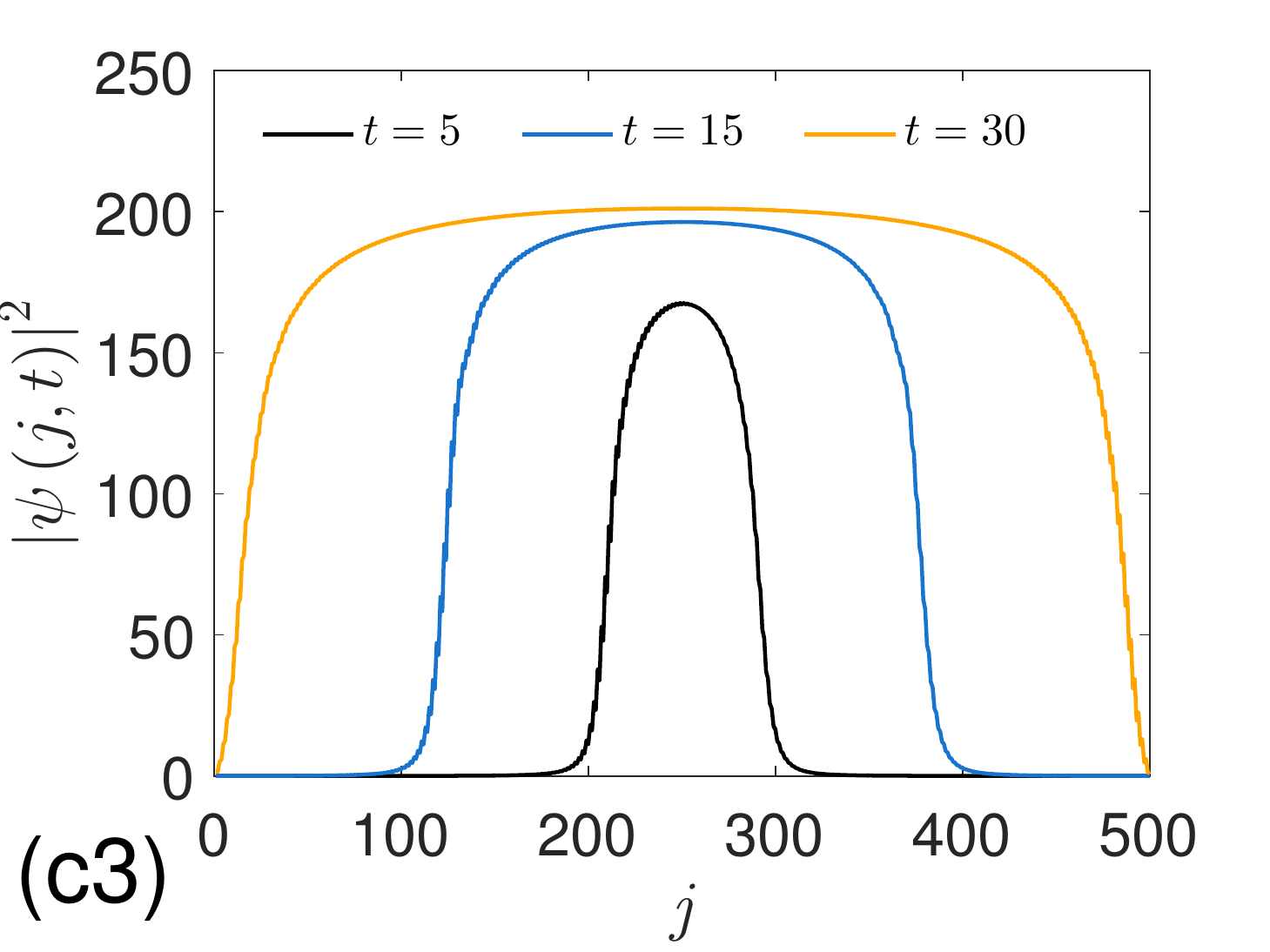}} %
\subfigure{\includegraphics[height=1.55in,width=1.76in]{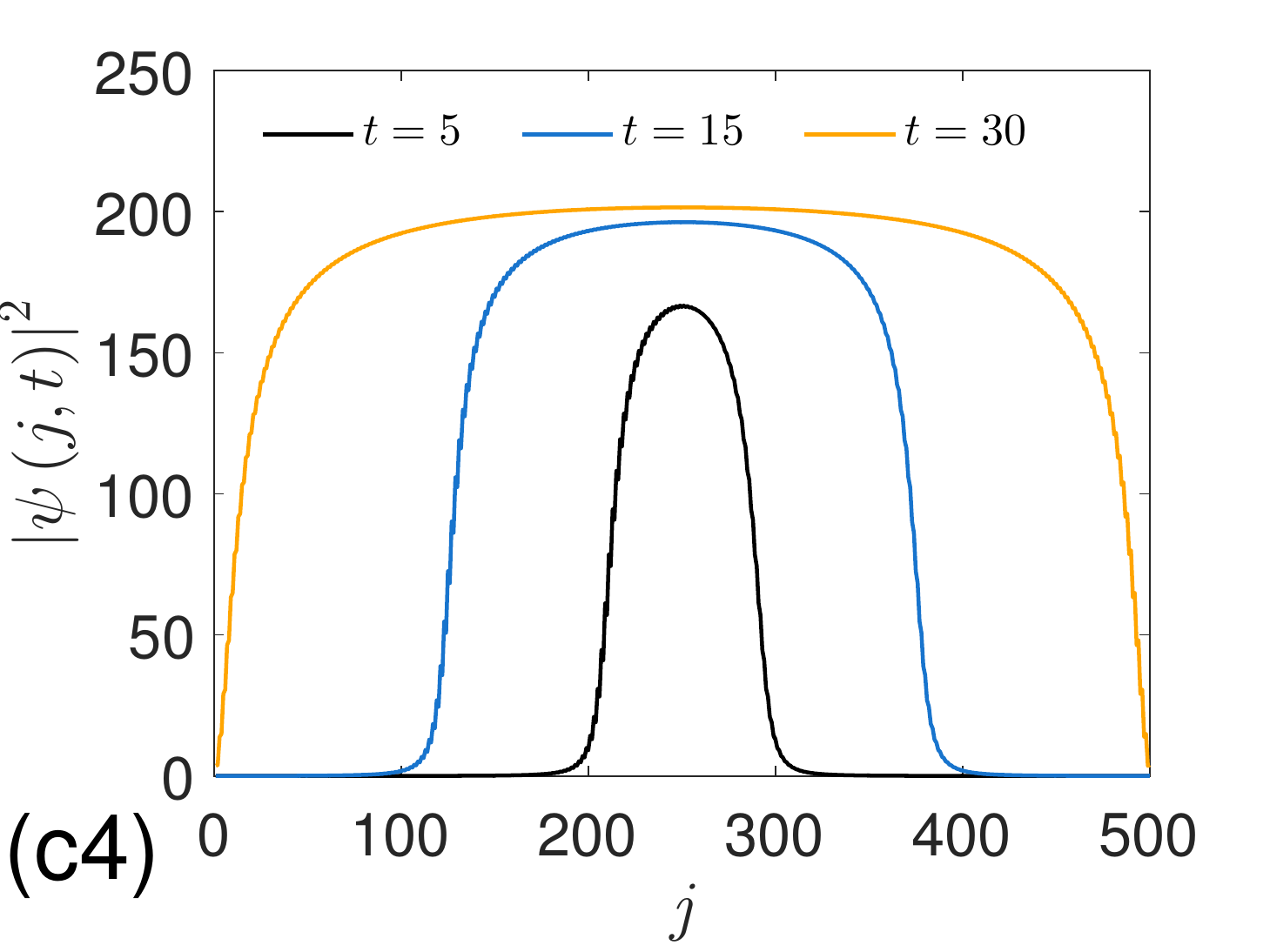}}
\caption{(Color online) Profiles of evolved wave packets for initial states
expressed \ in Eq. (\protect\ref{initial_state}) with $\protect\kappa _{0}=%
\protect\pi /2$\ and several typical $q$\ values. $q=0$ for (a1,a2,a3,a4), $%
q=0.02$ for (b1,b2,b3,b4), and $q=0.05$ for (c1,c2,c3,c4). (a1,b1,c1) Three-dimensional plots of evolved states for initial states plotted in (a2,b2,c2) obtained by numerical simulations. (a2,b2,c2) Plots of initial states from Eq. (\protect \ref{initial_state}). (a3,b3,c3) Profiles of evolved states at several instants obtained by numerical simulations. (a4,b4,c4) The same as (a3,b3,c3) but obtained by analytical expression in Eq. (\protect\ref{analytical_state}). The parameters for the SSH chain are $2N=500$, $\protect\delta =0.9$ and $\protect\gamma =1.8$. The time is in units of $\ 10{J}^{-1} $, where $J$ is the scale of the Hamiltonian and we take $J=1$. The analytical expressions accord with the numerical results well, especially for non-zero $q$. It shows that the evolved states are flat-top (rectangular shape for zero $q$) with the uniformly increasing width, exhibiting stationary lasing dynamics before touching the boundary.}
\label{Fig3}
\end{figure*}

\section{Lasing dynamics}

\label{Lasing dynamics}

\begin{figure*}[t]
\centering
\subfigure{\includegraphics[height=2.2in,width=2.32in]{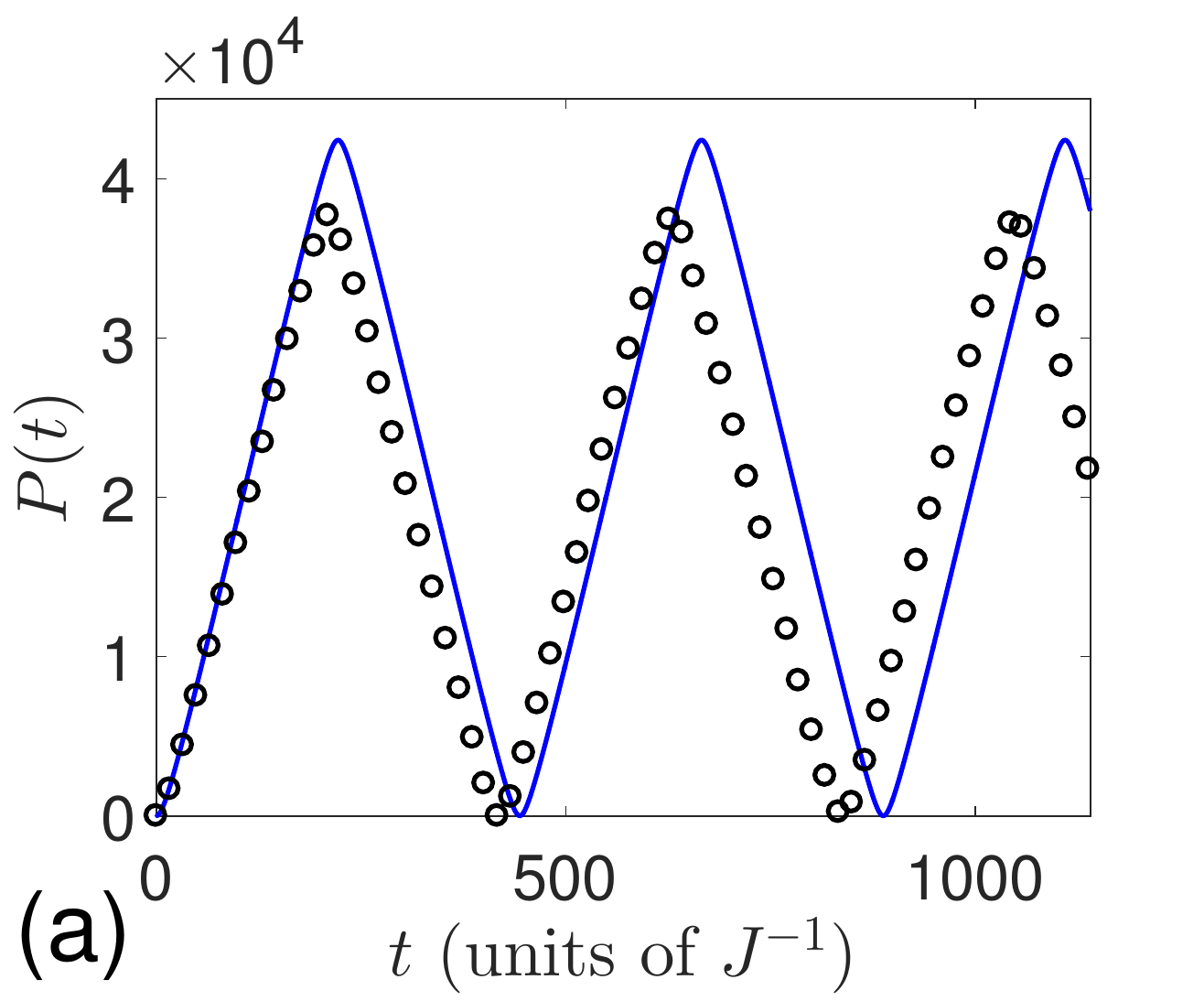}} %
\subfigure{\includegraphics[height=2.2in,width=2.32in]{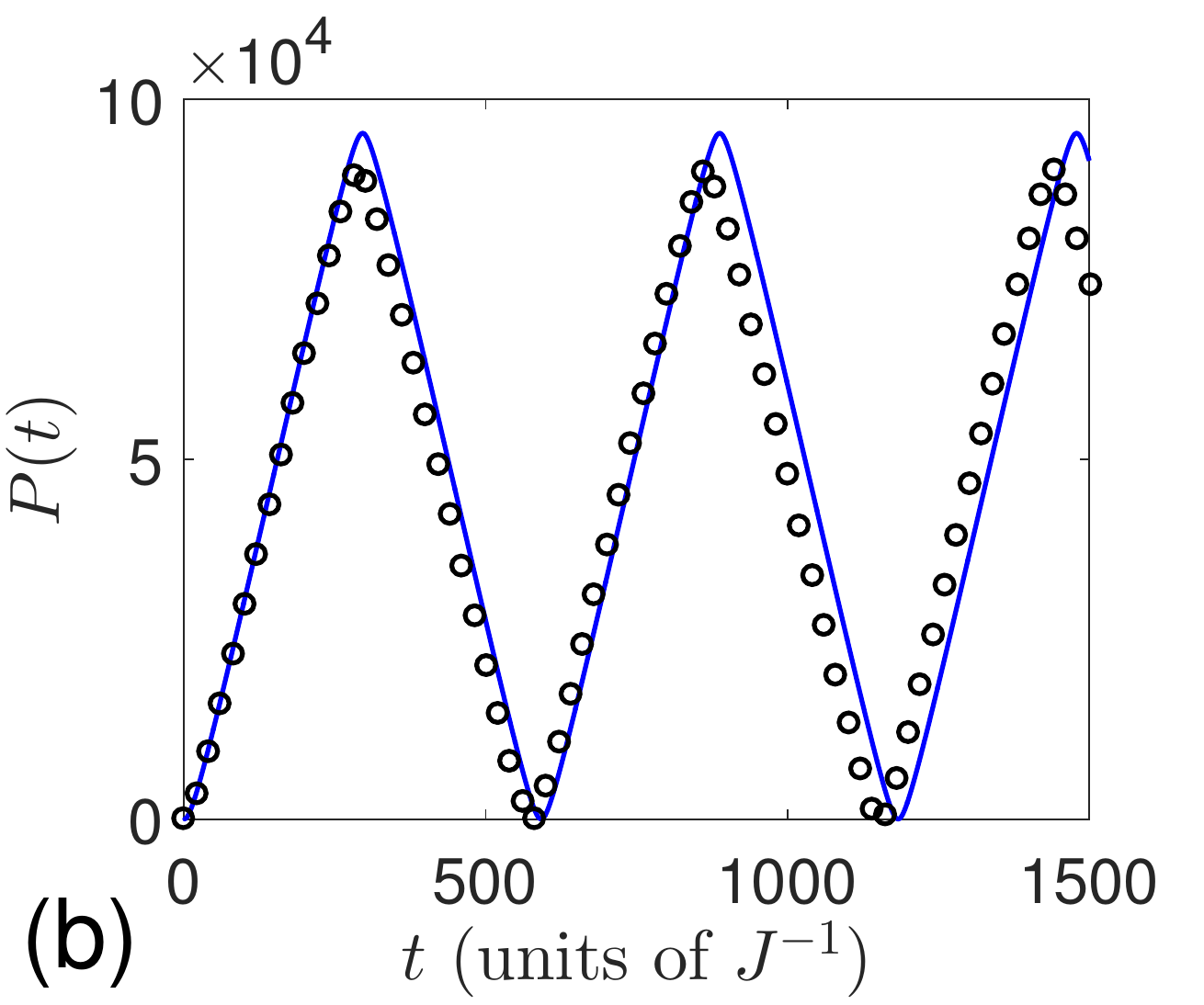}} %
\subfigure{\includegraphics[height=2.2in,width=2.32in]{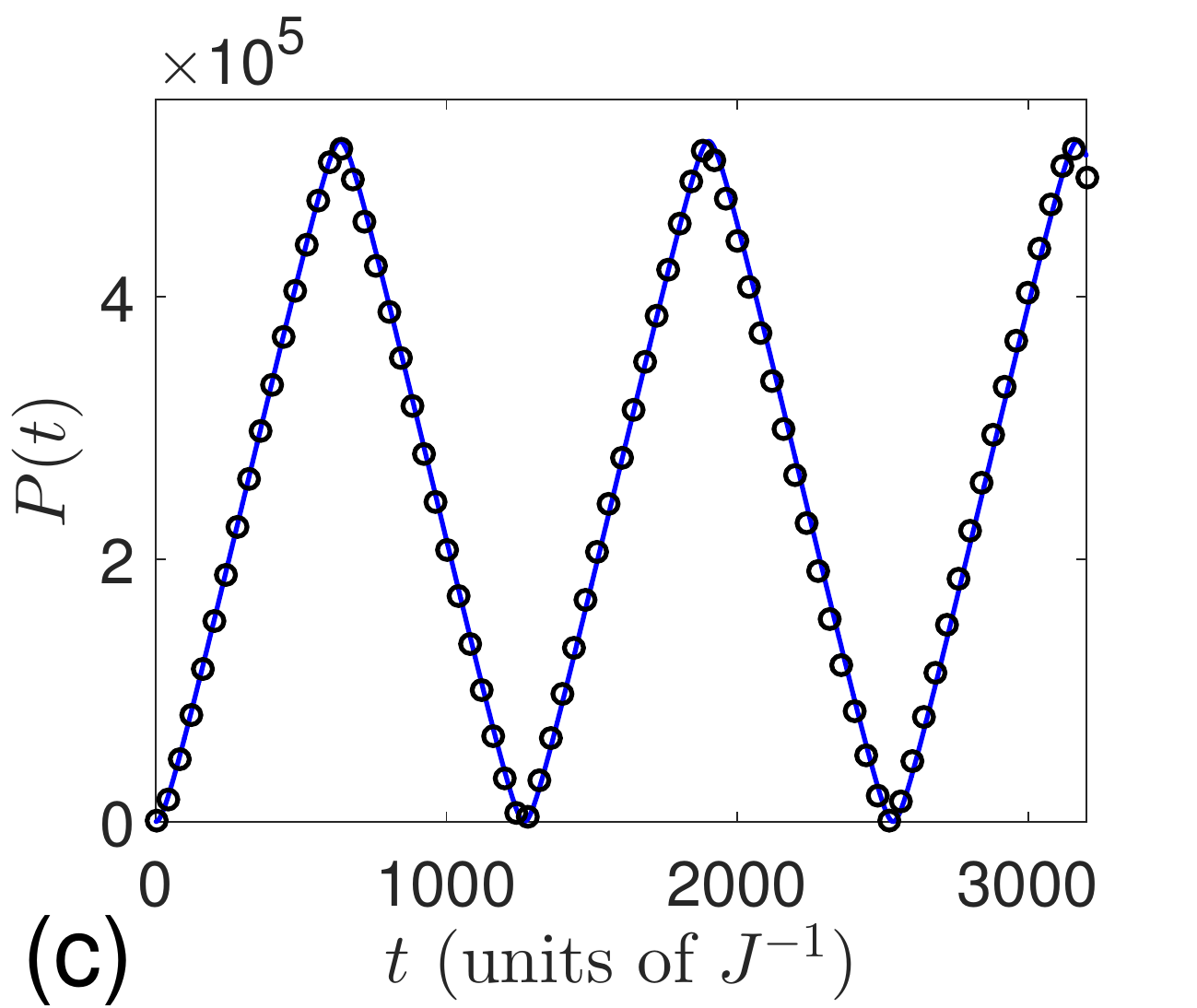}}
\caption{(Color online) Plots of $P\left( t\right) $ obtained by numerical
simulations (empty circle) and analytical expression (solid line) in Eq. (%
\protect\ref{P_t}). The parameters for the SSH chain are $2N=500$, $\protect%
\gamma =2\protect\delta$, and (a) $\protect\delta =0.8$, (b) $\protect\delta %
=0.9$, (c) $\protect\delta =0.98$, respectively. The initial states are
taken the form in Eq. (\protect\ref{initial_state}) with $\protect\kappa %
_{0}=\protect\pi /2 $ and $q=0.05$. It shows that $P\left( t\right) $ is a
triangle wave with period $\protect\tau =2\left( N+1\right) /\protect\sqrt{2%
\protect\delta (1-\protect\delta )}$. The analytical expressions accord with
the numerical results well, especially for the cases with strong
dimerization.}
\label{Fig4}
\end{figure*}

\begin{figure}[tbp]
\centering
\includegraphics[height=3.2in,width=3.5in]{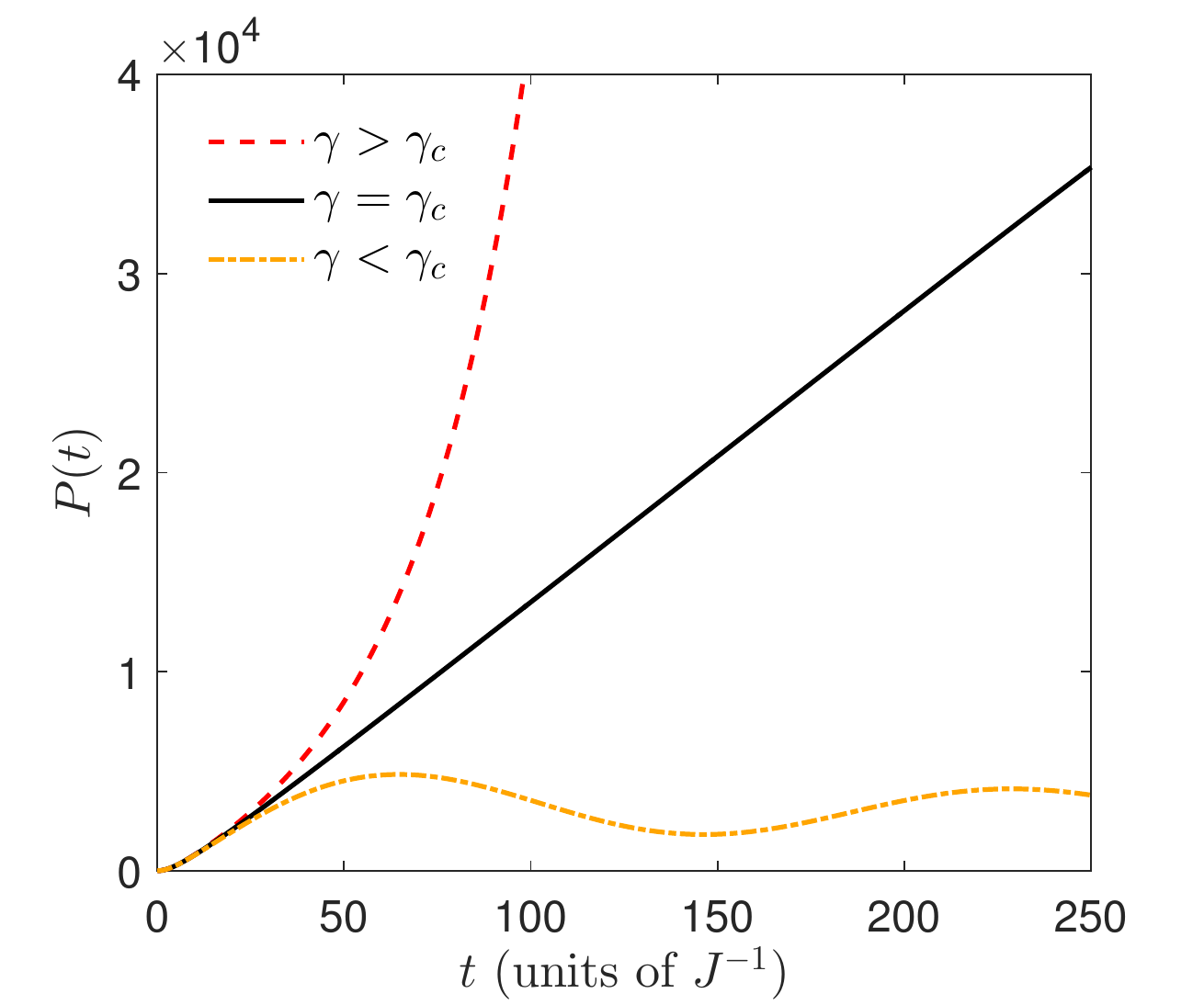} 
\caption{(Color online) Plots of $P\left( t\right) $ obtained by numerical
simulations for three typical values of $\protect\gamma $. The parameters
for the SSH chain are $2N=500$, $\protect\delta =0.9$ and $\protect\gamma _{c}=1.8$. The initial states are taken the form in Eq. (\protect\ref{initial_state}) with $\protect\kappa _{0}=\protect\pi /2$, and $q=0.02$. We
can see that $P\left( t\right) $\ is exponential for $\protect\gamma >\protect\gamma _{c}$, linear for $\protect\gamma =\protect\gamma _{c}$, and
oscillation for $\protect\gamma <\protect\gamma _{c}$, respectively. It
indicates that $\protect\gamma =\protect\gamma _{c}$ serves as the threshold
of the active laser medium. \ } \label{Fig5}
\end{figure}

In this section, we will investigate the dynamics for a class of specific
state. We focus on the initial state satisfying three conditions: (i) The
Dirac probability distribution is localized in the coordinate space; (ii)
The coefficient $c_{n}^{\sigma }$\ is non-vanishing only for small $n$;
(iii) It has $CT$\ symmetry.\ It is expected that such kinds of initial
state may possess lasing dynamics due to the following reasons: Although the
non-Hermitian SSH chain does not has EP at $\gamma _{c}$, its corresponding
ring system has a EP at zero energy. However, a local wave packet does not
know whether boundary condition is open or periodic unless its final state
touches the boundary. Then the wave packet should partially exhibit the EP
dynamics which obeys the time evolution in a Jordan block \cite%
{GraefePRL,GraefePRA,MoiseyevEP11}, when the initial state $\left\vert \psi
\left( 0\right) \right\rangle $\ has component of the coalescing eigenstates
of the SSH ring at EP. On the other hand, the above condition (ii) ensures
that $\left\vert \psi \left( 0\right) \right\rangle $\ may probably contain
such the coalescing component.\ From the analysis of last section,
conditions of (ii) and (iii) constrain the evolved state to possess
symmetric probability distribution and time-reflection symmetry, which make
the dynamics more convenient to describe.

Before focusing on specific initial states, we study some features of time
evolution for an initial state satisfying the above condition (ii). For
small $n$, we have%
\begin{equation}
\left\vert \psi \left( m\tau \right) \right\rangle =\sum_{n=1,\sigma =\pm
}c_{n}^{\sigma }\exp \left( -i2nm\sigma \pi \right) \left\vert \psi
_{n}^{\sigma }\right\rangle =\left\vert \psi \left( 0\right) \right\rangle ,
\end{equation}%
where 
\begin{equation}
\tau =\frac{2\pi }{\omega }=\frac{2\left( N+1\right) }{\sqrt{2\delta
(1-\delta )}},
\end{equation}%
and $m$\ is an integer. It indicates that the initial state $\left\vert \psi
\left( m\tau \right) \right\rangle $\ revivals periodically with period $%
\tau $. This property is not the direct result from the symmetry of the
system. In addition, for the present model, we have%
\begin{eqnarray}
\left\vert \psi \left[ \left( m+\frac{1}{2}\right) \tau \right]
\right\rangle &=&\sum_{n=1,\sigma =\pm }c_{n}^{\sigma }\exp \left( -in\sigma
\pi \right) \left\vert \psi _{n}^{\sigma }\right\rangle  \notag \\
&=&\sum_{n=1,\sigma =\pm }c_{n}^{\sigma }\mathcal{PT}\left\vert \psi
_{n}^{\sigma }\right\rangle ,
\end{eqnarray}%
based on the Eq. (\ref{PT-n}). Specifically, for a class of initial states
with a set of real (or total imaginary) $\left\{ c_{n}^{\sigma }\right\} $,
we have%
\begin{equation}
\left\vert \psi \left[ \left( m+\frac{1}{2}\right) \tau \right]
\right\rangle =\mathcal{PT}\left\vert \psi \left( 0\right) \right\rangle ,
\end{equation}%
i.e., such states revival periodically at the symmetric position with period 
$\tau /2$ \cite{ZXZ}. On the other hand, the dynamics of a Jordan block
should exhibit increasing probability with power law \cite%
{GraefePRL,GraefePRA,MoiseyevEP11}. The combination of the two results gives
us the following statement. In general, the probability should experience
both an increasing and decreasing processes within the time scale $\tau /2$.
It accords with the prediction of dynamics with time-reflection symmetry.

Now we construct a class of initial states which meet the above three
conditions. We will show that such states exhibit lasing dynamics during the
time evolution. The initial state has the form%
\begin{equation}
c_{n}^{\sigma }=\sigma \Lambda \sin \left( n\kappa _{0}\right) \frac{\exp
\left( -qn\right) }{n},
\end{equation}%
where $\Lambda $\ is normalization constant, $q\geqslant 0$,\ and $\kappa
_{0}\in \left( 0,\pi \right) $ are related to the shape and position of the
initial state, respectively. Obviously, $c_{n}^{\sigma }$\ is real and
vanishing for large $n$. We note that the initial state%
\begin{equation}
\left\vert \psi \left( 0\right) \right\rangle =\Lambda \sum_{n=1,\sigma =\pm
}^{N}\sigma \sin \left( n\kappa _{0}\right) \frac{\exp \left( -qn\right) }{n}%
\left\vert \psi _{n}^{\sigma }\right\rangle ,  \label{initial_state}
\end{equation}%
has both $\mathcal{PT}$\ and $\mathcal{CT}$\ symmetries for the case with $%
\kappa _{0}=\pi /2$. To demonstrate the localization of $\left\vert \psi
\left( 0\right) \right\rangle $\ in the coordinate space, we plot the
profile of $\left\vert \psi \left( 0\right) \right\rangle $\ with several
typical $\kappa _{0}$\ in Fig. \ref{Fig2}. It indicates that $\left\vert
\psi \left( 0\right) \right\rangle $\ is a local wave packet with the center
position $2N(\kappa _{0}/\pi )$.\ The time evolution of such local initial
is independent of the initial position within a certain time scale.\ Then we
can focus on the state $\left\vert \psi \left( 0\right) \right\rangle $\
with $\kappa _{0}=\pi /2$. The evolved state always has symmetric profile in
real space. Such kind of symmetric dynamics is convenient for analytical
analysis and the obtained result can be applied to the case with $\kappa
_{0}\neq \pi /2$\ by a simple translation due to the locality of the evolved
state.

To estimate the profile of the evolved state, we derive the evolved wave
vector\ in the following compact form 
\begin{eqnarray}
\left\vert \psi \left( t\right) \right\rangle &\approx &\Lambda
_{N}\sum_{j=1}^{N}\sum_{\rho ,\upsilon ,\eta =\pm 1}\arctan \left( \frac{%
\sin \theta }{e^{q}-\cos \theta }\right)  \notag \\
&&\times \left( -1\right) ^{j}\rho \upsilon \eta e^{i\eta \pi /4}\left\vert
2j-\frac{1}{2}\left( 1-\eta \right) \right\rangle ,  \label{analytical_state}
\end{eqnarray}%
where $\Lambda _{N}=\frac{\Lambda }{2}\sqrt{\left( -1\right) ^{N}/\left(
N+1\right) }$ and%
\begin{equation}
\theta =\rho \kappa _{0}+\frac{\upsilon \pi j}{N+1}+\omega t-\frac{\eta }{%
4\delta }.
\end{equation}%
In Fig. \ref{Fig3}, the profiles of $\left\vert \langle l\left\vert \psi
\left( t\right) \right\rangle \right\vert ^{2}$\ are plotted, which are
obtained by numerical simulations and approximate analytical expression in
Eq. (\ref{analytical_state}). It shows that the evolved state is a flat-top
wave packet with uniformly increasing width. After bouncing from the two ends
of the chain, it turns back to the initial state and starts the next
cycling. This observation can be explained by the following analysis for
special case.

When $q=0$, the expression of the evolved state reduces to%
\begin{eqnarray}
\left\vert \psi \left( t\right) \right\rangle  &\approx &\frac{\Lambda _{N}}{%
2}\sum_{j=1}^{N}\sum_{\rho ,\upsilon ,\eta =\pm 1}\left( -1\right) ^{j}\rho
\upsilon \eta e^{i\eta \pi /4}r\left( \theta \right)   \notag \\
&&\times \left\vert 2j-\frac{1}{2}\left( 1-\eta \right) \right\rangle ,
\end{eqnarray}%
where $r\left( x\right) $ is a periodic triangular function defined as: 
\begin{equation}
r\left( x\right) =\left( \pi -x\right) /2+n\pi ,x\in \lbrack 0,2\pi )+2\pi n,
\end{equation}%
with $n\in Z$. For $\kappa _{0}=\pi /2$, the Dirac norm of the evolved state
is%
\begin{eqnarray}
P\left( t\right)  &=&\left\langle \psi \left( t\right) \right\vert \psi
\left( t\right) \rangle   \notag \\
&\approx &-\frac{\Lambda ^{2}e^{-2q}}{2}\mathrm{Re}\left[ e^{-i2\omega
t}\Phi \left( e^{-4\left( q+i\omega t\right) },2,\frac{1}{2}\right) \right] 
\notag \\
&&+\Lambda ^{2}\sum_{\sigma =\pm }\sigma \mathrm{Li}_{2}\left( \sigma
e^{-2q}\right) ,  \label{P_t}
\end{eqnarray}%
where $\Phi \left( z,s,\alpha \right) =\sum_{n=0}^{\infty }z^{n}/\left(
n+\alpha \right) ^{s}$ is the Lerch transcendental function and $\mathrm{Li}%
_{n}\left( z\right) =\sum_{k=1}^{\infty }z^{k}/k^{n}$ is the Polylogarithm
function. In particular, taking $q=0$, the Dirac norm $P\left( t\right) $\
becomes a triangular wave 
\begin{equation}
P\left( t\right) \approx \frac{2\Lambda ^{2}\pi ^{2}}{\tau }\left\{ 
\begin{array}{cc}
t-n\tau /2, & t\in \lbrack 0,\tau /4)+n\tau /2 \\ 
-t+\left( n+1\right) \tau /2, & t\in \lbrack \tau /4,\tau /2)+n\tau /2%
\end{array}%
\right. ,  \label{P(t) q=0}
\end{equation}%
with $n\in Z$. As expected, it is the direct results of the uniform
expanding flat-top wave packet. We also plot the function $P\left( t\right) $
from Eq. (\ref{P_t}) and numerical simulation for several typical values of $%
\delta $ in Fig. \ref{Fig4}, which indicates that our analytical result
accords the numerical result well, especially for strong dimerization.

Obviously, such a stationary solution is related to the condition $\gamma
=\gamma _{c}$. It is natural to ask what happens when $\gamma $ deviates
from $\gamma _{c}$. To answer this question, numerical simulations are
performed by exact diagonalization. We plot the total probability as
function of time for three cases in Fig. \ref{Fig5}. It shows that the plot
is exponential, linear, and oscillation for $\gamma >\gamma _{c}$, $\gamma
=\gamma _{c}$, and $\gamma <\gamma _{c}$, respectively. It indicates that $%
\gamma _{c}$\ is the threshold for the lasing medium.

It is presumable that the lasing dynamics is independent of the position of
the initial state if the chain is large enough. It differ from the lasing
mechanics based on the SS in a non-Hermitian scattering center, in which
only the scattering center acts as the active lasing medium. The underlying
mechanism is the translational symmetry and the existence of EP in the bulk
region of the chain.

\section{Probability preserving and elastic collision}

\label{Probability preserving and elastic collision}

\begin{figure}[tbp]
\centering
\subfigure{
	\begin{minipage}[b]{0.48\linewidth}
		\includegraphics[height=1.55in,width=1.65in]{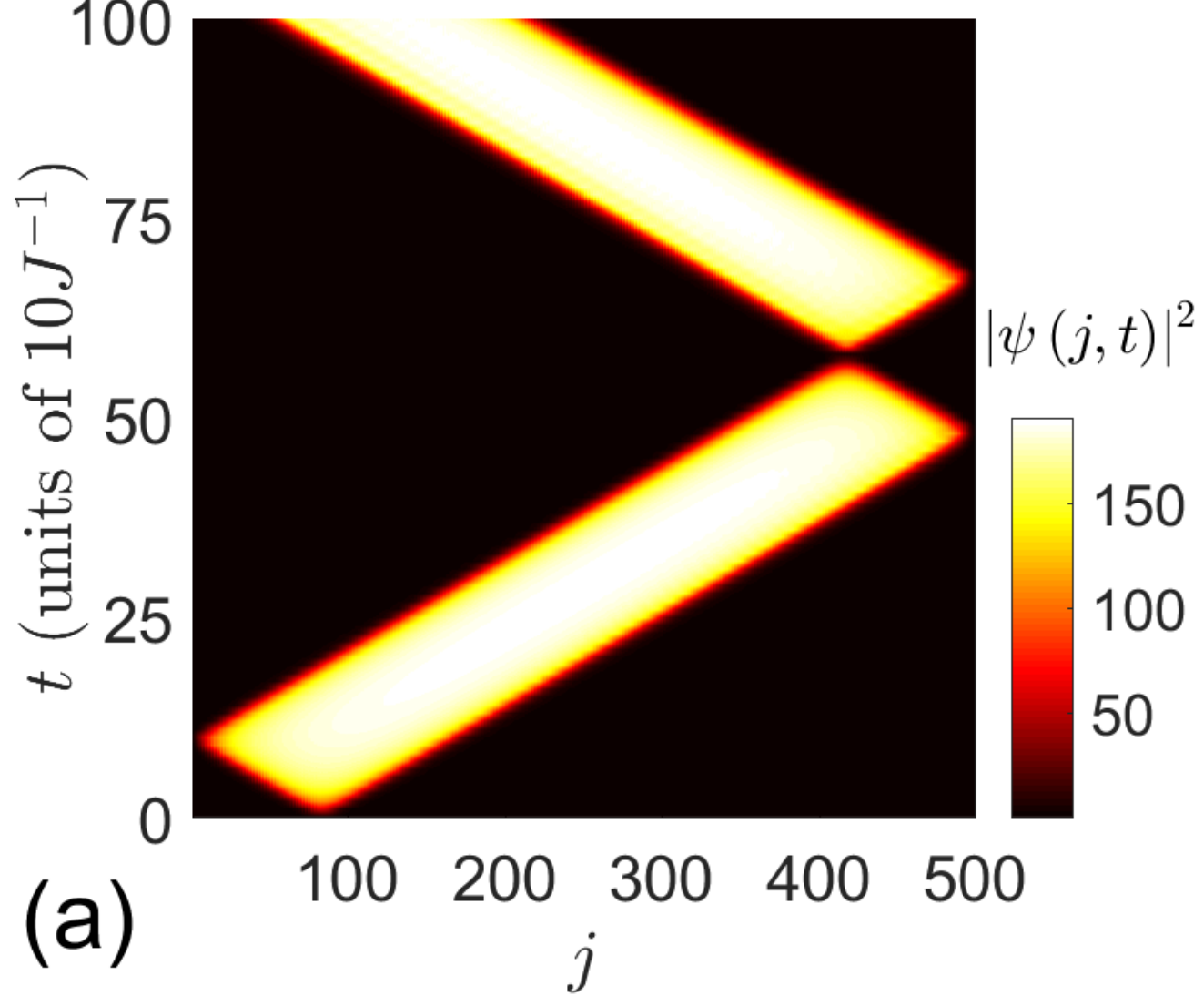}\vspace{4pt}
		\includegraphics[height=1.6in,width=1.65in]{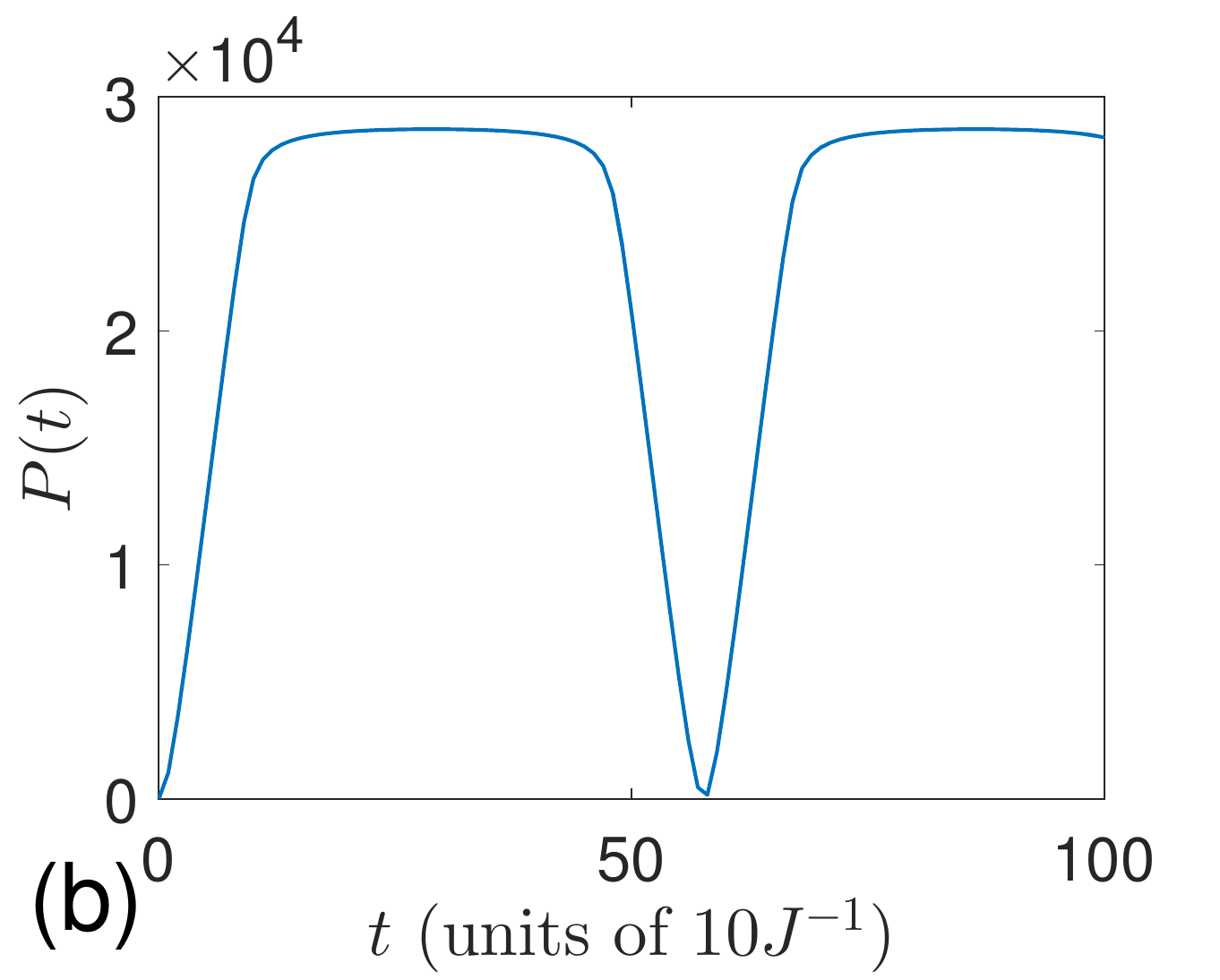}\vspace{2pt}
\end{minipage}} 
\subfigure{
	\begin{minipage}[b]{0.48\linewidth}
		\includegraphics[bb=0 20 390 627, height=3.4in,width=1.8in]{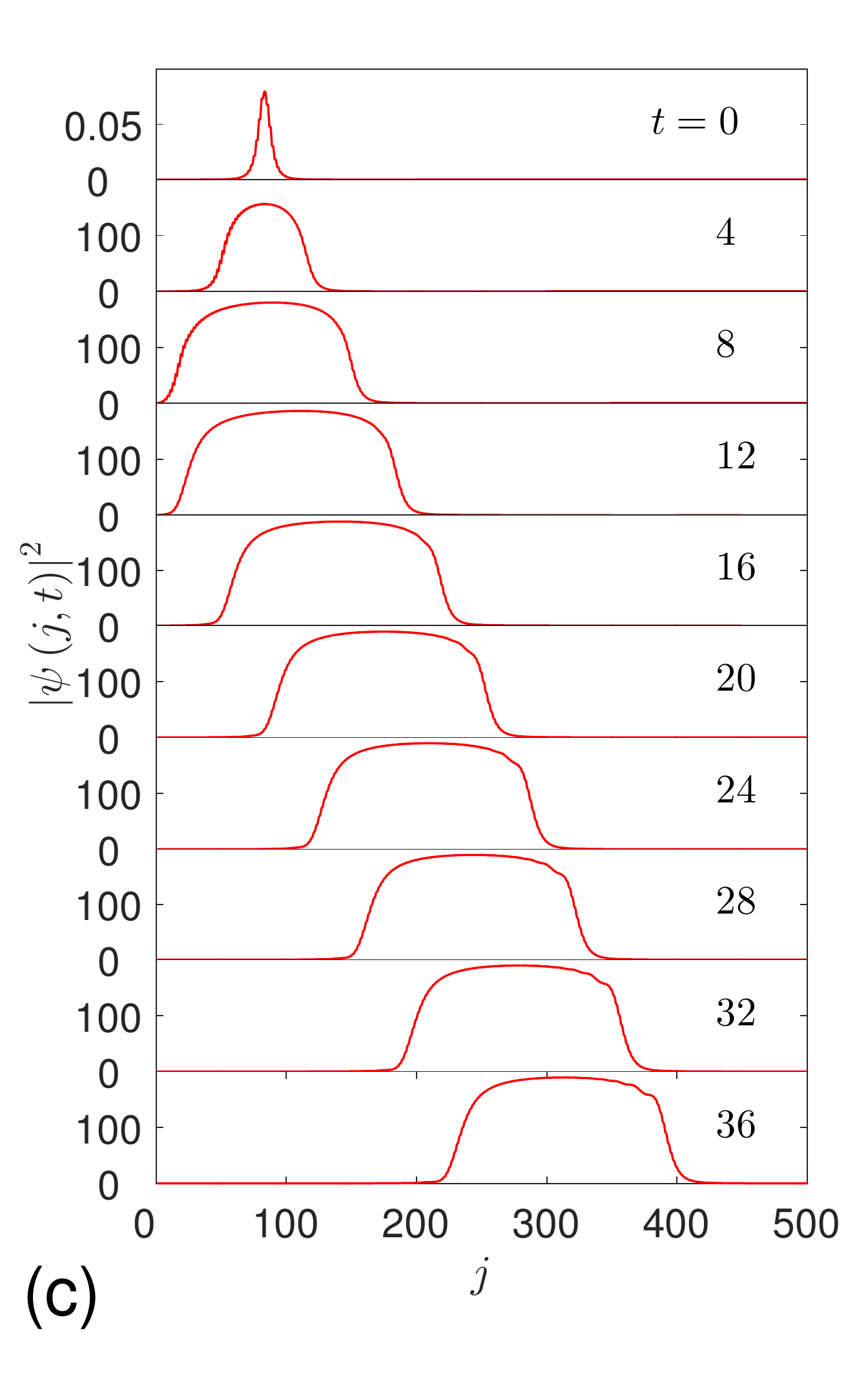}\vspace{2pt}	
\end{minipage}} 
\caption{(Color online) (a) Three-dimensional plots of evolved wave packets for initial states expressed in Eq. (\protect\ref{initial_state}) with $\protect\kappa _{0}=\protect\pi /6$ and $q=0.05$, obtained by numerical simulation. (b) Plot of $P\left( t\right) $ for the time evolutions for (a). (c) The profiles of the evolved wave packet at several typical instants. The parameters for the SSH chain are $2N=500$, $\protect\delta =0.9$ and $\protect\gamma =1.8$. It shows that the evolved states becomes a travelling wave packet after the first reflection on the one side, i.e., it propagates as a translational motion, preserving the probability before the subsequent reflection on the other side.} \label{Fig6}
\end{figure}

\begin{figure*}[t]
\centering
\subfigure{
	\begin{minipage}[b]{0.23\linewidth}
	\includegraphics[height=1.55in,width=1.65in]{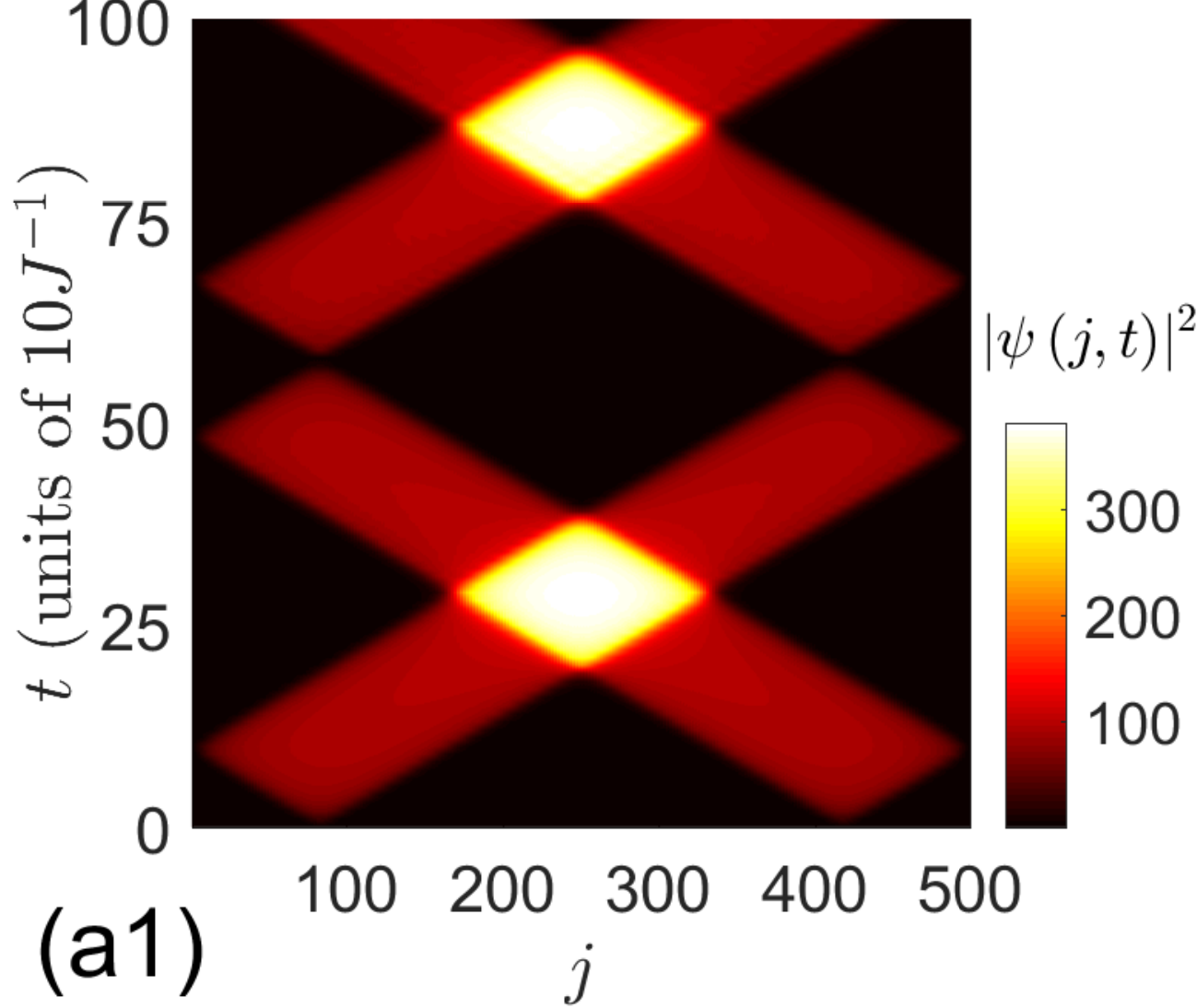}\vspace{4pt}
	\includegraphics[height=1.55in,width=1.65in]{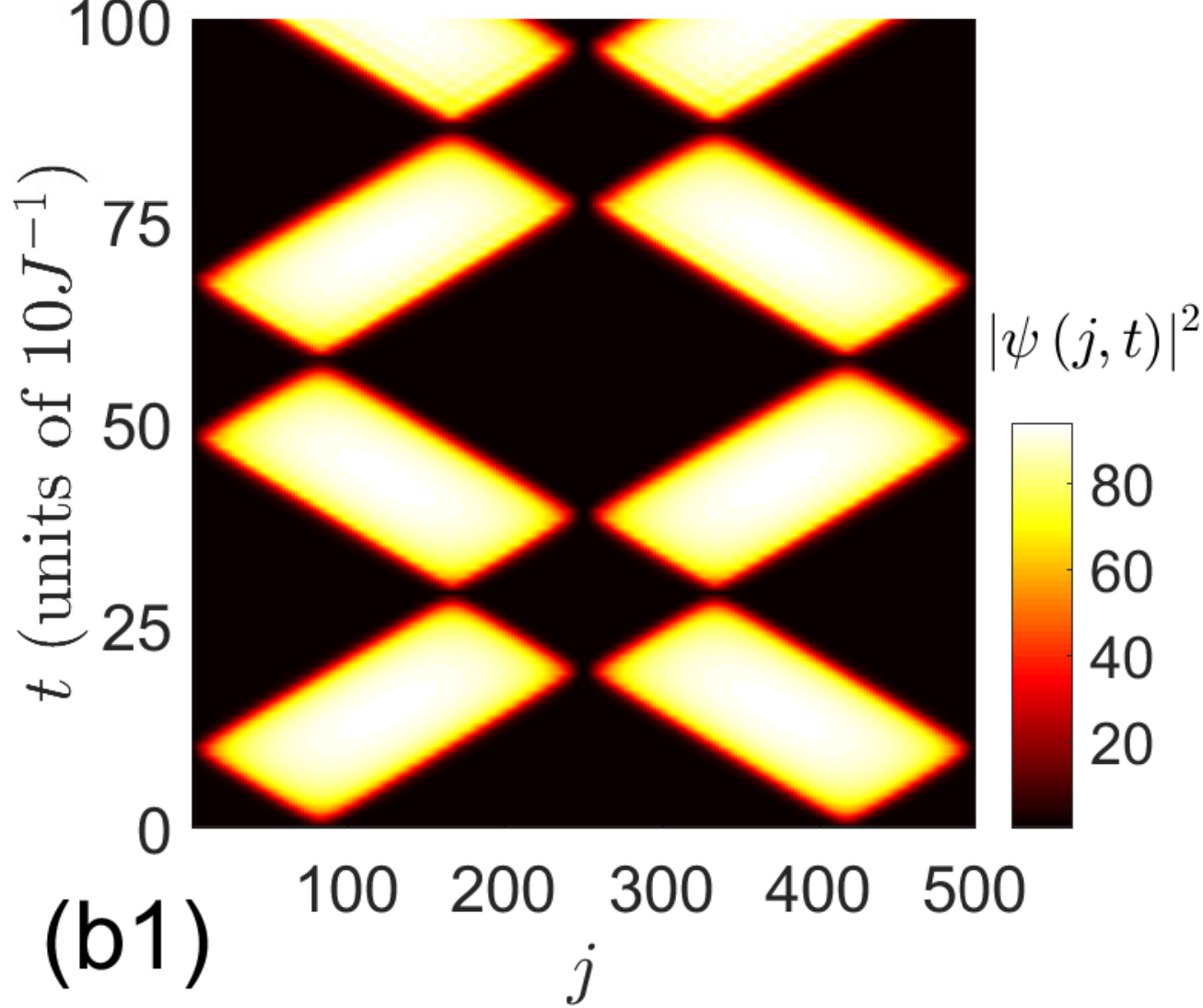}\vspace{2pt}	
    \end{minipage}} 
\subfigure{
	\begin{minipage}[b]{0.23\linewidth}
		\includegraphics[height=1.6in,width=1.65in]{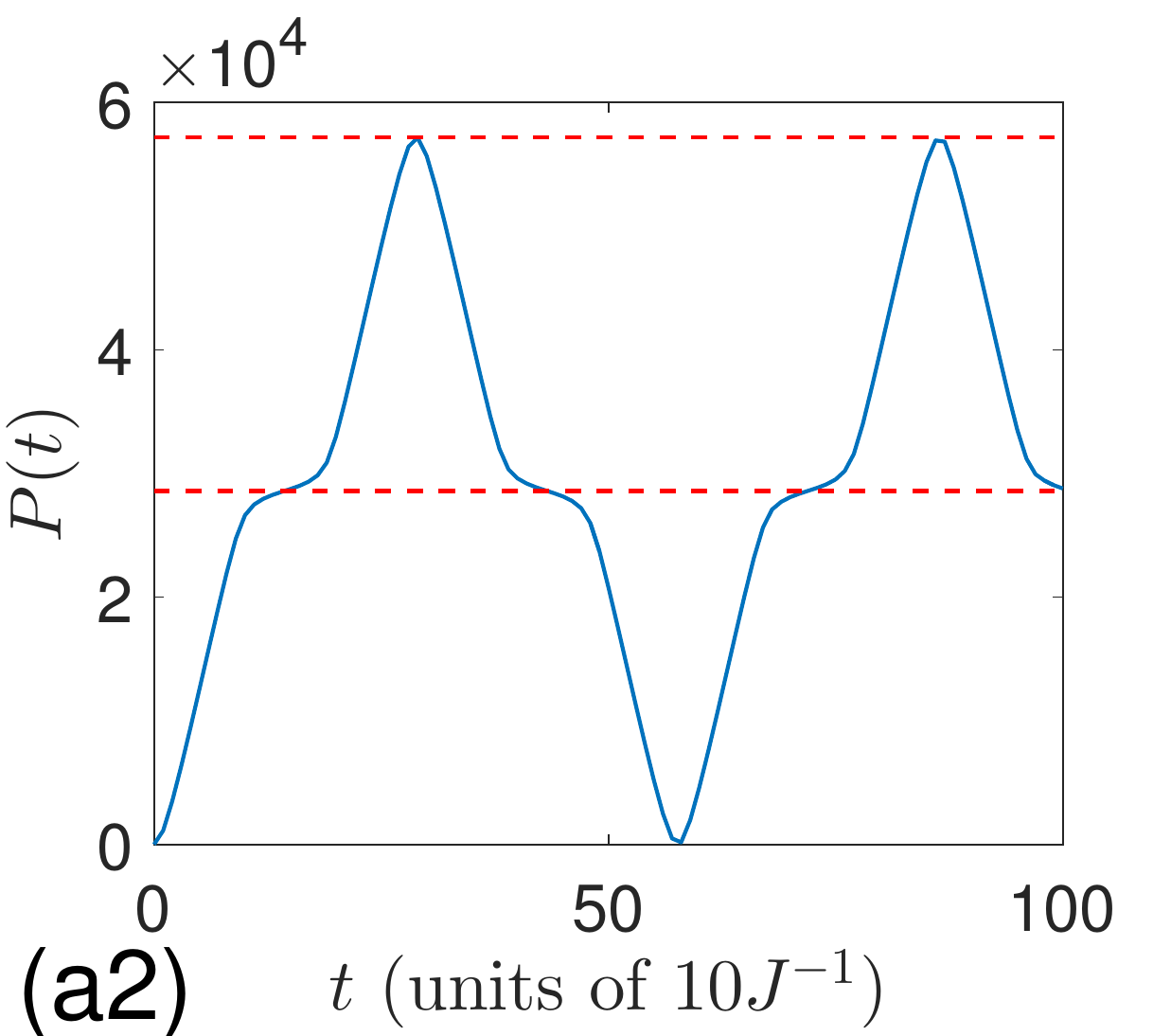}\vspace{4pt}
		\includegraphics[height=1.6in,width=1.65in]{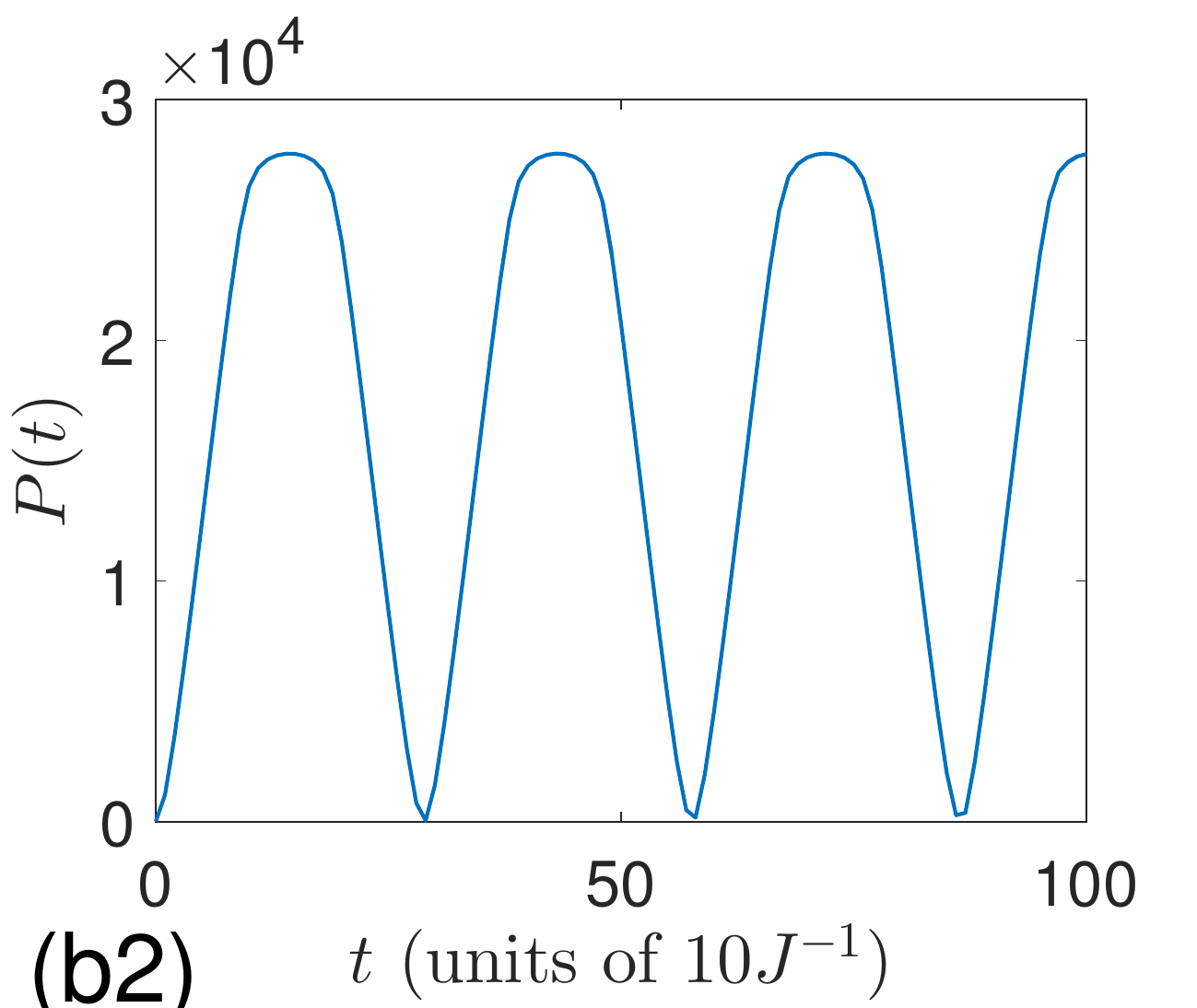}\vspace{2pt}	
    \end{minipage}} 
\subfigure{
	\begin{minipage}[b]{0.25\linewidth}
	\includegraphics[bb=0 20 367 627, height=3.4in,width=1.85in]{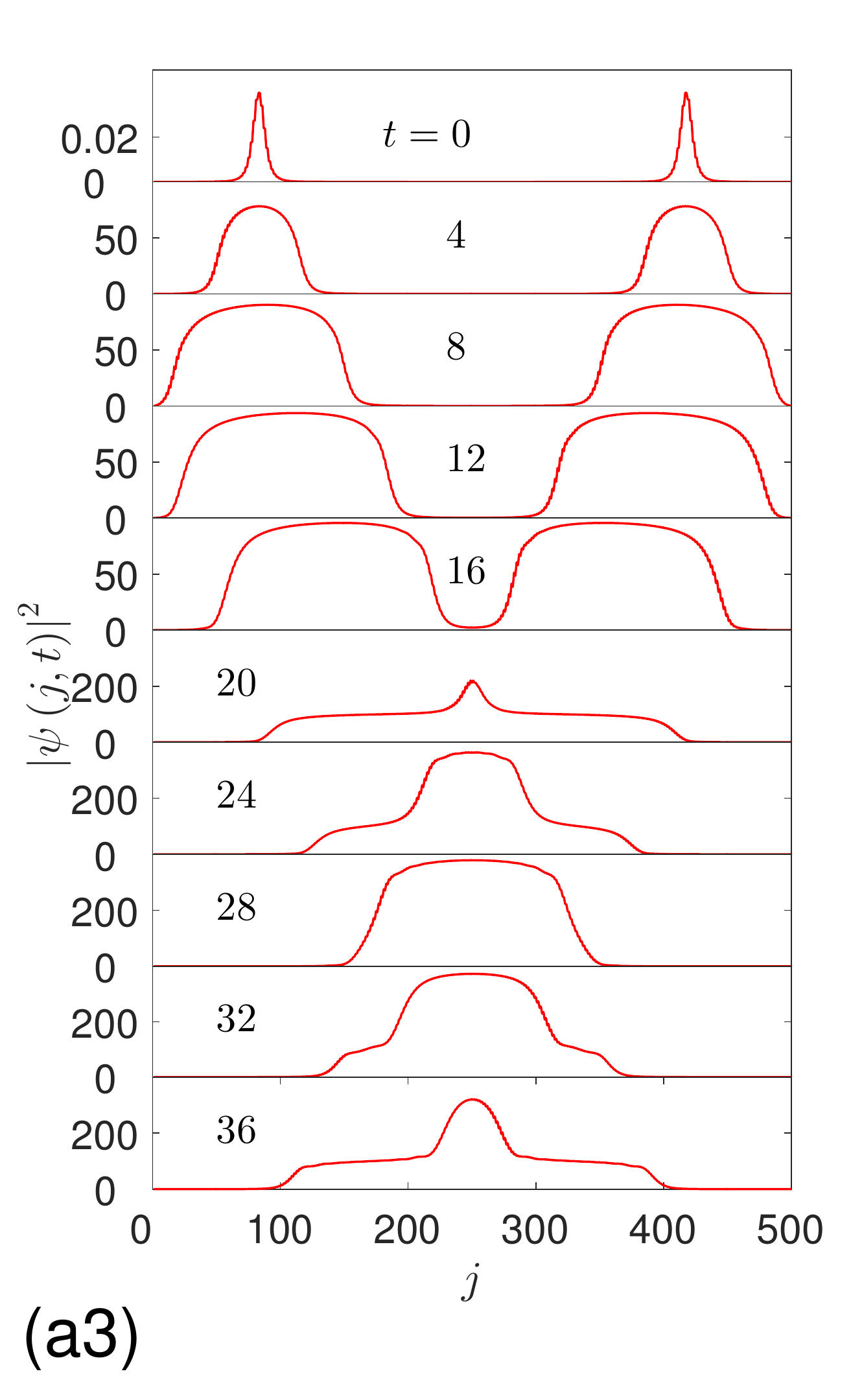}\vspace{2pt}	
    \end{minipage}} 
\subfigure{
	\begin{minipage}[b]{0.245\linewidth}
		\includegraphics[bb=0 20 367 620, height=3.4in,width=1.85in]{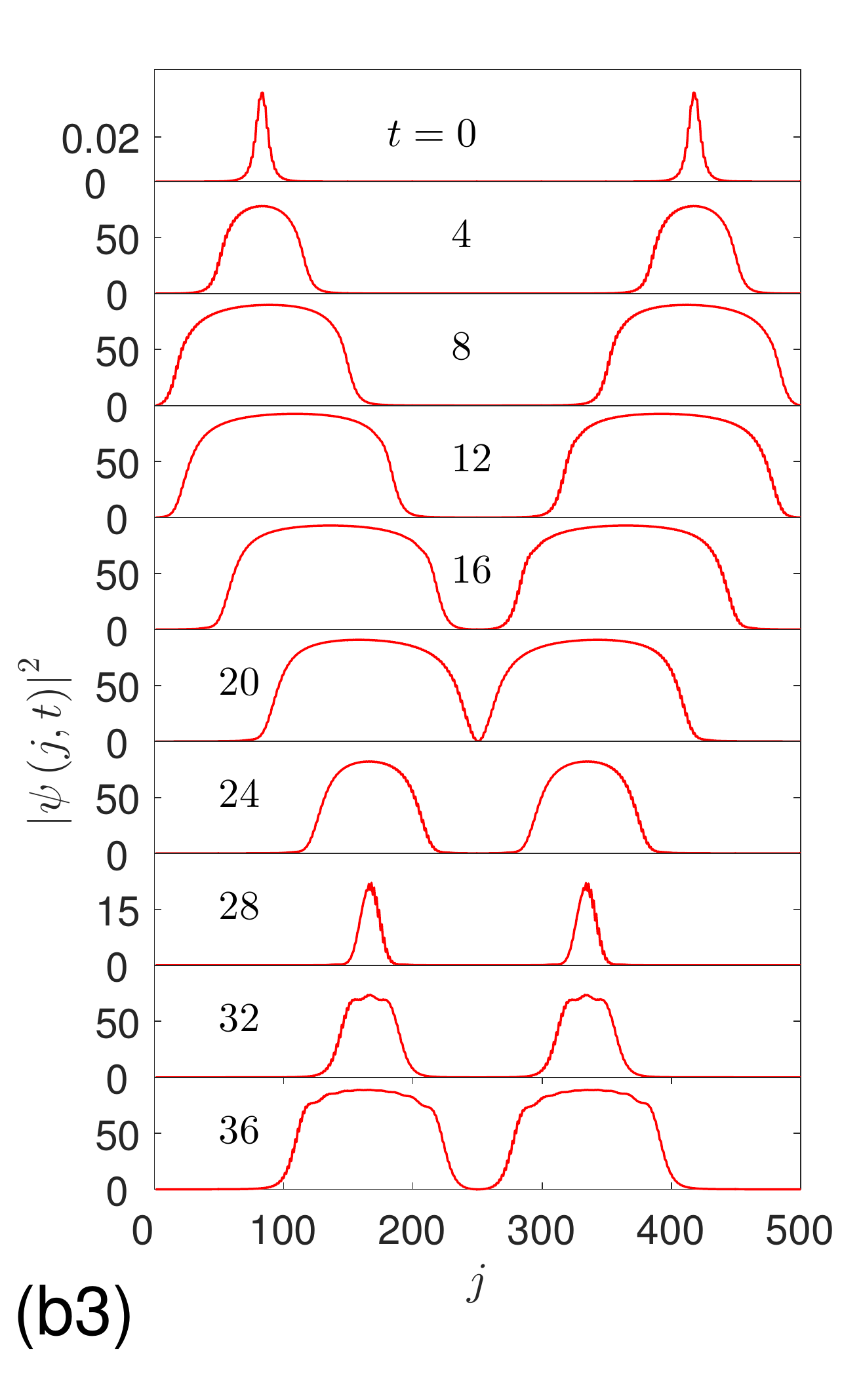}\vspace{2pt}	
    \end{minipage}}
\caption{(Color online) The same as that in Fig. \protect\ref{Fig6}, but for
initial states (a1) $\left\vert \Psi _{+}\right\rangle $ and (b1) $%
\left\vert \Psi {-}\right\rangle $ expressed in Eq. (\protect\ref{2WP}) with 
$\protect\kappa _{01}=\protect\pi /6,$ $\protect\kappa _{02}=5\protect\pi /6$%
, obtained by numerical simulation. (a2) and (b2) are plots of $P\left(
t\right) $ for the time evolutions for (a1) and (b1), respectively. (a3) and (b3) are the profiles of the evolved wave packets at several typical instants for (a1) and (b1), respectively. We see
that when two wave packets are separated, the total probability is always
constancy, while changes when they overlap. In the case of (a2), the
probability doubles as indicated by two dotted lines. It is due to the
anomalous interference phenomenon, in which there are no interference
patterns. In contrast, from (b2) we see that the probability turns to very
small, as if two wave packets annihilate together, or absorb each other. It
is a peculiar dynamics, no counterpart in Hermitian system.}
\label{Fig7}
\end{figure*}

From the analysis in last section, we find that the extending rectangular
wave or flat-top wave packet bounds back at two ends of the chain. We note
that there are two features in the reflection process: (i) There is no
interference pattern (standing wave) as usual. (ii) This process acts like a
time-refection one, i.e., the profiles of input and output are identical. In
this section, we will study the underlying mechanism of this phenomenon. For
this purpose, it will be convenient to think in terms of symmetric case and
extends the conclusion to a general case, i.e.,\ taking the case with $%
\kappa _{0}=\pi /2$. In this case, the initial state is superposition of
eigen states with small odd $n$. These eigen states are all long wave-length
standing wave, the superposition of which have no ability to form
interference pattern with small-frequency.

On the other hand, Fig. \ref{Fig3} and analytical analysis show that two
edges of the flat-top wave packet touch the two ends of the chain at instant $%
t=\tau /4$. We note that 
\begin{equation}
\left\vert \psi \left( \tau /4\right) \right\rangle =\sum_{n=1,\sigma =\pm
}c_{2n-1}^{\sigma }\exp \left( -in\sigma \pi \right) i\sigma \left\vert \psi
_{2n-1}^{\sigma }\right\rangle ,
\end{equation}%
which are both $\mathcal{PT}$\ and $\mathcal{CT}$\ symmetric, satisfying%
\begin{eqnarray}
\mathcal{PT}\left\vert \psi \left( \tau /4\right) \right\rangle
&=&\left\vert \psi \left( \tau /4\right) \right\rangle , \\
\mathcal{CT}\left\vert \psi \left( \tau /4\right) \right\rangle
&=&i\left\vert \psi \left( \tau /4\right) \right\rangle ,
\end{eqnarray}%
taking $\left\vert \psi \left( \tau /4\right) \right\rangle $ as initial
state, and using the conclusion (Eq. (\ref{time-reflection})) of Sec. \ref%
{Quasi-symmetric dynamics}, we obtain 
\begin{equation}
\left\vert \langle j\left\vert \psi \left( \tau /4+\Delta t\right)
\right\rangle \right\vert ^{2}\approx \left\vert \langle j\left\vert \psi
\left( \tau /4-\Delta t\right) \right\rangle \right\vert ^{2}
\end{equation}%
i.e., the reflection process is symmetric about the instant $t=\tau /4$. We
refer this phenomena as to elastic reflection due to the fact that it is
analog to a mass-spring system in classical physics.

On the other hand, the dynamics before reflection is the same for an initial
state located in anywhere of the chain. Then the elastic reflection also
happens for the initial state with $\kappa _{0}\neq \pi /2$. We will show
that the combination of two such features leads to a probability preserving
dynamics which usually appears in a Hermitian system. Such a process occurs
when consider an initial state locates far from the center of the chain. As
extension of the wave packet, one of its edges moves in opposite direction
after the elastic bounce from one end of the chain. Then two edges of the
wave packet move in the same direction with the same speed, which results in
a translational motion of the wave packet, preserving the Dirac probability.
Fig. \ref{Fig6} Plot of profile evolved state for initial state with $\kappa
_{0}\neq \pi /2$.

Now we consider the time evolution for initial state as a superposition of
two wave packets. Such an investigation is trivial for a Hermitian system.
However, some unexpected phenomena may be found in a non-Hermitian system
although it is also a linear system. This is because the Dirac probability
is not defined by a canonical inner product (biorthonormal inner product).
The initial state is taken in the form%
\begin{eqnarray}
\left\vert \Psi _{\pm }\right\rangle &=&\frac{\Lambda }{\sqrt{2}}%
\sum_{n=1,\sigma =\pm }\sigma \left[ \sin \left( n\kappa _{01}\right) \pm
\sin \left( n\kappa _{02}\right) \right]  \notag \\
&&\times \frac{\exp \left( -qn\right) }{n}\left\vert \psi _{n}^{\sigma
}\right\rangle ,  \label{2WP}
\end{eqnarray}%
where $\kappa _{01}$\ and $\kappa _{02}$\ determine the initial locations of
two wave packets. The trajectories and profiles of two wave packets are clear
when they do not overlap. We are interested in what happens when they meet
together. To answer this question, we employ the numerical simulation to
compute the probability of the evolved state. Fig. \ref{Fig7}, shows that
the dynamics of $\left\vert \Psi _{+}\right\rangle $\ behaves as a Hermitian
one, while $\left\vert \Psi _{-}\right\rangle $\ exhibits a peculiar
behavior: It looks like that two wave packets cancel each other out when they
meet together.

\section{Summary}

\label{Summary}

In summary, we have investigated the non-Hermitian analogue of an active
laser medium, and find a scenario for the mechanism of lasing in the
framework of quantum mechanics. We have proposed alternative lasing
mechanism induced by the EP in a finite non-Hermitian system rather than SS
in an infinite system with a non-Hermitian scattering center. The key
difference between two mechanisms is that a laser can be fired everywhere on
the former while only at the scattering center in the latter. The present
non-Hermitian system also exhibits many peculiar dynamic behaviors, such as
elastic reflection, collision and probability preserving translational
propagation, etc. The underlying mechanism of such features is\ the balance
of distortion and staggered imaginary potentials, while an SS laser solution
does not require the balance. In general, the position of EPs are a little
different for a same model but with different boundary conditions, or more
generally speaking, with and without some defects. When the initial state is
local in coordinate space, its dynamics within certain period of time is
independent of the boundary condition and its initial location. This
provides a way to explore the dynamics, which is the combination of a
quasi-Hermitian and Jordan block time evolutions in a general non-Hermitian
system. So the key point in practice is the specific class of initial
states. For the present model, the linear increase of probability arises
from such a combination. It can be seen from the nonzero overlap 
\begin{equation}
\left\vert \langle \phi _{c}\left\vert \psi \left( 0\right) \right\rangle
\right\vert \approx \frac{\sqrt{\delta \left( 1-\delta \right) }\Lambda }{%
N\delta }\arctan \left( \frac{\sin \kappa _{0}}{\sinh q}\right) ,
\end{equation}%
where 
\begin{equation}
\left\vert \phi _{c}\right\rangle =\frac{1}{\sqrt{2N}}\sum_{j=1}^{N}\left(
-1\right) ^{j}(\left\vert 2j-1\right\rangle +i\left\vert 2j\right\rangle ),
\end{equation}%
is the coalescing eigen vector of the conjugate Hamiltonian $H^{\dag }$ in
Eq. (\ref{H}) with periodic boundary condition. Our results, on the one
hand, provide an alternative lasing theory in the context of non-Hermitian
quantum mechanics, on the other hand, indicates non-Hermitian system is a
fertile ground for many unknown features in physics.

\acknowledgments This work was supported by National Natural Science Foundation of China (under Grant No. 11874225).

\end{document}